\documentclass[12pt]{article}
\usepackage{graphics}
\usepackage{amssymb,amsmath,amsfonts,epsfig,bm}
\usepackage{setspace}
\usepackage[sort&compress]{natbib}
\bibpunct{}{}{,}{s}{}{,}
\usepackage{caption}

\numberwithin{equation}{section}

\title{Exploration of Effective Potential Landscapes\\
  using Coarse Reverse Integration}

\author{
Thomas A. Frewen\thanks{Princeton University,
Department of Chemical Engineering,
Engineering Quadrangle,
Olden Street, Princeton, NJ 08544;
{\it e-mail: tfrewen@princeton.edu.}}
\and Gerhard Hummer\thanks{Laboratory of Chemical Physics, National Institute of Diabetes and
Digestive and Kidney Diseases, National Institutes of Health, Bethesda, MD 20892-0520;
{\it e-mail: Gerhard.Hummer@nih.gov.}}
\and Ioannis G. Kevrekidis\thanks{Princeton University,
Department of Chemical Engineering, PACM and Mathematics,
Engineering Quadrangle,
Olden Street, Princeton, NJ 08544;
{\it e-mail: yannis@princeton.edu.}}
}

\date{\today}
\begin{document}

\maketitle
\begin{abstract}
We describe a reverse integration approach for the exploration of
low-dimensional effective potential landscapes.
Coarse reverse integration initialized on a ring of coarse states
enables efficient ``navigation'' on the landscape terrain: escape from
local effective potential wells, detection of saddle points,
and identification of significant transition paths between wells.
%
%
We consider several distinct ring evolution modes: backward
stepping in time, solution arc--length, and effective potential.
The performance of these approaches is illustrated for a
deterministic problem where the energy landscape is known
explicitly.
Reverse ring integration is then applied to ``noisy'' problems
where the ring integration routine serves as an outer ``wrapper''
around a forward-in-time inner simulator.
Three versions of such inner simulators are considered:
a system of stochastic differential equations,
a Gillespie--type stochastic simulator, and a molecular dynamics simulator.
In these ``equation-free" computational illustrations, estimation techniques are
applied to the results of short bursts of ``inner'' simulation to
obtain the unavailable (in closed form) quantities (local drift and
diffusion coefficient estimates) required for reverse ring
integration; this naturally leads to approximations of the effective landscape.
\end{abstract}
\newpage
\doublespacing
\section{Introduction}
When an energy landscape perspective is applicable, the dynamics of a complex
system appear dominated by gradient-driven descent into energy wells,
occasional trapping in deep
minima, and  transitions between minima via passage over  saddle
points through thermal ``kicks''.
A paradigm for this landscape picture is the trapping of protein
configurations in metastable states en route to the dominant folded
state.
The underlying  energy landscape is often likened to a roughened funnel with trapped
states corresponding to local free energy minima \cite{Wolyn95}.

Important features on energy surfaces include local minima and their
associated basins of attraction, saddle points, and minimum energy paths (MEPs)
between neighboring minima passing through these saddles.
Besides the identification of such
landscape features, establishing the details of their connectivity is
a task of considerable importance.
Knowledge of the relative depths
of landscape minima provides thermodynamic information. The kinetics of transitions
between such states is determined by the type of ``terrain'' (smooth,
rugged, etc.) that surrounds and separates them, in particular by the location and height of the low-lying saddles.
The identification of  important low
energy molecular conformations in computational
chemistry \cite{Kolossvary96}, and determination of protein and peptide
folding pathways \cite{Brooks01}, to name but a few, rely on an
ability to perform intelligent, targeted searches of the energy
landscape.

Molecular dynamics (MD) and Monte Carlo (MC) simulations on energy
landscapes are typically limited in the time scales they can explore by the
difference between system thermal energy and the height of
transitional energy barriers.
A significant fraction of MD and MC
simulation time is spent ``bouncing around'' in local minima.
Energy barriers
separating minima cause this type of trapping and the
result is long waiting times between infrequent, but interesting, transition events.
An array of techniques have been proposed to
overcome such time scale limitations including bias-potential approaches
\cite{Wang01,Huber94}, accelerated dynamics \cite{Voter02},
coarse-variable dynamics \cite{Gear02,Hummer03}, and transition
path sampling \cite{Dellago98,Dellago98a},
allowing extensive exploration of the energy surface and its
transition states.
The adaptive bias force method \cite{Darve01,Chipot05}
efficiently samples configurational space in the presence of high
free energy barriers via estimation of the force exerted on the system along
a well-defined reaction coordinate.
Short bursts of appropriately
initialized simulations are used in coarse-variable dynamics
\cite{Hummer03} to infer the deterministic and  stochastic components
of motion parametrized by an appropriate set
of coarse variables.
The use of a history-dependent bias potential in Ref.~13 ensures
that minima are not revisited, allowing for efficient exploration
of a free energy surface parametrized by a few coarse coordinates.
Accelerated dynamics methods such as
hyperdynamics and parallel replica dynamics \cite{Voter02}
``stimulate'' system trajectories trapped in local
minima to find appropriate escape paths while preserving the relative
escape probabilities to different states.
Transition path sampling \cite{Dellago98} generalizes importance
sampling to trajectory space and requires no prior knowledge of a
suitable reaction coordinate (see also transition interface sampling\cite{vanE05}).

Many energy landscape search methods have been devised (too numerous
to discuss in detail here). ``Single-ended'' search approaches
(where only the initial state is known) are based on eigenvector-following (mode-following)
\cite{Baker86,Taylor85,Cerjan81,Poppinger75,Kolossvary96,Goto98} and
have been used to refine details of minimum energy paths (MEPs) close to saddle points
\cite{Page88,Gonzalez91}; methods purely for efficient saddle point
identification also exist \cite{Ionon93,Miron01,Henkelman99}.
\emph{Chain-of-state} methods are a more recent class of double-ended
 searches that evolve a chain of replicas (system states or  ``images''),
distributed between initial and final
states, in a concerted manner\cite{Jonsson98}. The original elastic band method
 \cite{Elber87,Gillilan92} has been refined and extended many times\cite{Henkelman00,Trygubenko04}.
More recently string methods \cite{E02,E05,Peters04}, which evolve
smooth curves with intrinsic parametrization, have been used to
locate significant paths between two states. The \emph{Global
Terrain} approach of Lucia and coworkers\cite{Lucia04,Lucia02}
exploits the inherent \emph{connectedness} of  stationary points
along valleys and ridges on the landscape for their systematic
identification.

We build here on the ``equation-free'' formalism of Ref.~35 whose
purpose is to enable the
performance of macroscopic tasks using appropriately designed
computational experiments with microscopic models.
The approach focuses on systems for which the coarse-grained, effective evolution equations
are assumed to exist but are not available in closed form.
One example is the case of ``legacy'' or ``black-box'' codes,
dynamic simulators which, given initial conditions, integrate forward in time
over an interval $\Delta t$.
Alternatively, the effective evolution equation for the system may
be the unknown closure of a microscopic simulation model such as
kinetic MC or MD.
Rico-Martinez et al.\cite{RicoM04} have used reverse integration
in conjunction with microscopic {\em forward-in-time} simulators
to access reverse time behavior of coarse variables (see also
Ref.~37).
%
Hummer and Kevrekidis \cite{Hummer03} used coarse reverse
integration to trace a {\em one-dimensional} effective free energy
surface (and to escape from its minima) for alanine dipeptide in
water.
In this paper we use
reverse integration in two dimensions:
a ring of system initial states is evolved (forward in time in the ``inner" simulation,
and then reverse in the ``outer", coarse integration) to explore
{\em two-dimensional} potential energy (and, ultimately, free energy) surfaces.
The ring is evolved along the component of the local energy gradient
(projected normal to the ring) while a nodal redistribution
scheme is used that slides nodes along the ring so that they remain equidistributed
in ring arclength, ensuring efficient sampling.
Transformation of the
independent variable in our basic ring evolution equation
results in several distinct {\em stepping modes}.

The paper is organized as follows.
%
In Section \ref{sect:ringmethod} we present our reverse ring integration approach.
Ring evolution equations are developed with time, arc-length, or
(effective) potential energy as
the independent variable.
We illustrate these stepping modes for a deterministic
problem with a smooth energy landscape (M\"{u}ller-Brown potential).
In Section \ref{sect:exampleprobs} reverse ring integration is
investigated for three ``noisy" problems: a system of stochastic
differential equations (SDEs), a Gillespie--type stochastic
simulation, and a molecular dynamics simulation of a protein
fragment in water.
Estimates of the
quantities in the ring evolution equation are found by data processing of
the results of appropriately initialized short bursts of the
``black-box'' inner simulator.
The extension to stepping in free energy is discussed.
We conclude with a brief discussion of the results and of the potential
extension of the approach to more than two coarse dimensions.

\section{Reverse integration on energy surfaces}
\label{sect:ringmethod}

Here we present a method for (low-dimensional) landscape exploration
motivated by reverse projective integration, on the one hand, and
by algorithms for the computation of (low-dimensional) stable manifolds
of dynamical systems on the other.
Reverse projective integration \cite{Gear04a}
uses short bursts of {\em forward in time} simulation
of a dynamical system to estimate a local time derivative
of the system variables, which is then used to take
a large {\em reverse} projective time step via polynomial
extrapolation.
This type of computation is intended for problems with a large
separation between many fast stable modes and a few (stable or
unstable) slow ones; the long-term dynamics of the problem will then
lie on an attracting, invariant ``slow (sub)manifold".
Reverse projective integration allows us to compute ``in the past'',
approximating solutions {\em on this slow manifold} by only using
the forward-in-time simulation code.

%
After each reverse projective step the reverse solution will be
slightly ``off manifold" (see Figure~1);
the initial part of the next short forward
burst will then bring the solution back close to the manifold, while the
latter part of the burst will provide the time derivative estimate
necessary for the next backward step.
One clearly does not integrate the {\em full system} backward in time (the
fast stable modes make this problem very ill-conditioned);
 it is the
slow ``on manifold" backward dynamics that we attempt to follow.
The approach can be used for deterministic dynamical systems of the
type described; however, it was developed having in mind
problems arising in atomistic/stochastic simulation where the
dynamic simulator is a molecular dynamics or kinetic Monte Carlo code.
If the dynamics
can be well described by a (low-dimensional) effective ODE, or
 an effective SDE, characterized by
a potential (and by a (low-dimensional) effective free energy surface),
 reverse projective
integration can be implemented as an ``outer" algorithm, wrapped around
the high-dimensional ``inner" deterministic/stochastic simulator.
The combination of short bursts of fine scale ``inner" forward in time
simulation with data processing and estimation and
then with coarse-grained ``outer" reverse integration
can then be used to systematically explore these effective potentials
(and associated effective free-energy surfaces).

A natural set of protocols for such an exploration has already
been developed (in the deterministic case) in dynamical systems
theory -- indeed, algorithms for the computation of
low-dimensional {\em stable} manifolds of vectorfields provide the
``wrappers" in our context (see the review in Ref.~38).
%
This is easily seen in the context of a two-dimensional gradient dynamical
system: an isolated local minimum of the associated potential is a stable
fixed point and, locally, the entire plane is its stable manifold; the
potential is a function of the points on this plane.
In our 2-dimensional case, we approximate this stable manifold by
a linearization in the neighborhood of the fixed point -- this
could be in the form of a ring of points surrounding the fixed
point.
One can then integrate the gradient vectorfield forward or backward in time
(see Figure~2)
keeping track
of the evolution of this initial ring; using the gradient nature of the system,
one can compute, as a byproduct, the potential profile.
%

%
%
Various versions of such ``reverse ring integration" have been previously used
for visualizing two-dimensional stable manifolds of vector
fields.
Johnson et al. \cite{Johnson97} evolved a ring stepping in space-time arclength
(see below) with empirical mesh adaptation and occasional addition of nodes to preserve ring
resolution, building up a picture  of the manifold as the ring
expands.
Guckenheimer and Worfolk \cite{Guckenheimer93} used algorithms based upon
geodesic curve construction to evolve a circle of points according to
the underlying vectorfield.
A survey of methods for the computation of (un)stable manifods of
vectorfields can be found in Ref.~38, including approaches for the
approximation of $k$-dimensional manifolds.
In this paper we will restrict ourselves to the two-dimensional case
(and thus, eventually, explore two-dimensional effective free energy surfaces).

Clearly, forward integration of our ring constructed based on local linearization
around an isolated minimum, will generate a sequence of shrinking rings
converging to the minimum (stable fixed point).
For a two-dimensional gradient vectorfield, backward (reverse) ring integration
will ``grow" the ring -- and as it grows on the plane, the potential on the
ring evolves ``uphill" in the initial well, possibly toward unstable
(saddle-type) stationary points.

A critical issue in tracking the reverse evolution of such a ring
is its distortion, as different portions of it evolve with different
rates along the ``stable manifold" (here, the plane).
Dealing with the distortion of this closed curve and the deformation of
an initially equidistributed mesh of discretization points on it requires
careful consideration; similar problems arise, and are elegantly dealt
with, in (forward in time) computations with the string method \cite{E02,Ren03}.
While we will first implement our reverse ring integration on a deterministic
gradient problem (for descriptive clarity), our aim is to use it as a wrapper
around atomistic/stochastic inner forward-in-time simulators; three such
illustrations will follow.
\subsection{The deterministic two-dimensional case}
\label{sect:example2d}
Consider a simple, two-dimensional gradient system of the form
\begin{equation}
\frac{d\bm {x}}{dt}=\left[\begin{array}{c} dx/dt \\
    dy/dt\end{array}\right]=-\nabla V(x,y).
\label{eq:gradientsys}
\end{equation}
In this case, since the vectorfield is explicitly available, with
$\bm{x}$ in $\mathbb{R}^2$, we can
perform reverse integration by simply reversing the sign of the
right hand side of equation~(\ref{eq:gradientsys}); reverse {\em projective}
integration will only become necessary in cases where the (effective)
 potential is not known, and the corresponding gradients
need to be estimated from forward runs of a many-degree-of-freedom
atomistic/stochastic simulator.
Note also that here the dependent variables $x$ and $y$ are known
(as are the corresponding evolution equations).
For high dimensional problems with a low-dimensional {\em effective}
description, selection of such good reduced variables (observables)
is nontrivial; we will briefly return to this in the
Discussion.

We start with a simple illustrative example: the M\"{u}ller-Brown potential
energy surface \cite{Mulle79}, which is often used to evaluate landscape search
methods since the minimum energy path (MEP) between its minima
deviates significantly from the chord between them.
We focus here on reverse ring evolution starting around a local
minimum in the landscape and approaching the closest saddle point as
the ring samples the well.

The potential is given by
\begin{equation}
 V(x,y)=\sum_{i=1}^4 A_i
\exp\left[a_i(x-x_i^0)^2+b_i(x-x_i^0)(y-y_i^0)+c_i(y-y_i^0)^2\right]
\end{equation}
where $A=(-200,-100,-170,15)$, $a=(-1,-1,-6.5,0.7)$, $b=(0,0,11,0.6)$,
\\ $c=(-10,-10,-6.5,0.7)$, $x^0=(1,0,-0.5,-1)$, and $y^0=(0,0.5,1.5,1)$.
The neighborhood of the M\"{u}ller-Brown potential we explore is shown
in Figure~3
along with a listing of the fixed
points, their energy, and their classification.
We first discuss the initialization of the ring, and then three
different forms of ``backward stepping": time-stepping,
arclength-stepping in (phase space)$\times$(time) and
potential-stepping.
Our initial ring will be the $V=-105$ energy contour surrounding the
minimum at $(0.62,0.03)$.

%
%
%
%
%
A ring is a smooth curve ${\bm\Phi}$, here in two dimensions. In our
implementation, we discretize
this curve and denote the instantaneous position of the discretized ring by the
vectors $\Phi_i\equiv{\bm\Phi}(\alpha_i,t)=\left[x(\alpha_i,t), y(\alpha_i,t)\right]$
(with $\Phi_i$ in $\mathbb{R}^2$, $\alpha_i$ in $\mathbb{R}$)
for the coordinates of the $i^{th}$ discretization node, where
$\alpha_i$ is a suitable parametrization.
A natural choice is the normalized arc-length along the ring with
$\alpha_i\in[0,1]$, as in the string method, but now with periodic
boundary conditions.
Note that one does not need to initialize on an exact isopotential
contour; keeping the analogy with local stable manifolds of a
dynamical system fixed point, one can use the local linearization
-- and more generally, local Taylor series -- to approximate a closed
curve on the manifold.
Anticipating the ``energy-stepping" reverse evolution mode, however,
we start with an isopotential contour here.
This requires an initial {\em point} on the surface; we then trace the
isopotential contour
passing through this point using a scheme which resembles the
sliding stage in the ``Step and Slide'' method of Miron and
Fichthorn \cite{Miron01} for saddle point identification.
We simply ``slide" along the contour to generate a curve ${\bm\Gamma}$, moving (in some pseudo-time
$\tau$) perpendicular to the
local energy gradient according to
\begin{equation}
\frac{d{\bm\Gamma}}{d\tau}=\left[\begin{array}{c} \partial V/\partial y\\-\partial V/\partial x\end{array}\right].
\label{eq:ringinit}
\end{equation}
 Points along the curve ${\bm\Gamma}$ provide initial conditions for
 ring nodes. Figure~4a
 illustrates ring initialization
 starting ``in the well", close to the isolated local minimum, resulting in a closed ring.

We note that our approach is closely related to established landscape search
techniques based on following Hessian
eigenvectors \cite{Baker86,Taylor85,Cerjan81,Poppinger75,Kolossvary96,Goto98};
here the computation
is performed in a dynamical systems setting: we use a dynamic simulator to
estimate time derivatives (and through them local potential gradients) on demand.

\subsection{Modes of Reverse Ring Evolution}
\label{sect:ringmodes}
\subsubsection{Time Stepping}
When every point on a curve evolves backward in time, it makes sense
to consider the evolution of the entire curve in the direction of
the component of the energy gradient normal to it (as also happens
for forward time evolution in ``string'' methods, commonly used to
identify minimum energy paths (MEPs)\cite{E02}).
Ring nodal evolution is given by
\begin{equation} \frac{d\Phi_i}{dt}=-(\nabla V(\Phi_i))^\perp+r\hat{T}
\label{eq:ringevolve0}
\end{equation}
where $\hat{T}$  is the unit tangent vector to ${\bm\Phi}$ at
$\Phi_i$, with
$\hat{T}=\frac{\partial\bm\Phi}{\partial\alpha}/|\frac{\partial\bm\Phi}{\partial\alpha}|$
evaluated at $\Phi_i$,
 and $\emph{r}$ is a Lagrange multiplier field \cite{E02} (determined by the choice of ring
parametrization) used to distribute nodes evenly along the ring.
The component of potential gradient normal to the ring $(\nabla
V)^\perp$ is defined as follows
\begin{equation}
(\nabla V)^\perp=\nabla
V-(\nabla V\cdot\hat{T})\hat{T}=\nabla V-(\nabla V)^\parallel
\end{equation}
where $(\nabla V)^\parallel$ is the component of the gradient
parallel to the ring.
For the general case where $(\nabla V(\Phi_i))^\perp$ is unavailable in
closed form (e.g.\ the inner integrator is a black-box timestepper)
we use (multiple short replica) simulations for each discretization
node on the ring to estimate it, as will be discussed in Section \ref{sect:exampleprobs}.
%
%
In practice, the tasks of node stepping and redistribution are often
split into separate stages.
The  term involving $\emph{r}$ in equation~(\ref{eq:ringevolve0}) is
first omitted, and nodal stepping is performed solving, backward in
time, the $N$-node spatially discretized form
\begin{equation} \frac{d\Phi_i}{dt}=F(\Phi_i,t)=-\nabla
  V(\Phi_i)^\perp, \quad i=1,2,\ldots,N
\label{eq:ringevolveNODE}
\end{equation}
where $\Phi_i$ denotes the position of node $i$ in the
discretized ring.
The normalized arc-length coordinate $\alpha_i$ associated with the
$\emph{i}^{th}$ node is approximated using the linear distance
formula \vspace{0.25cm}
\begin{equation}
 \alpha_i=\frac{\sum_{j=1}^i{\sqrt{(\Phi^x_j-\Phi^x_{j-1})^2+(\Phi^y_j-\Phi^y_{j-1})^2}}}{\sum_{j=1}^N{\sqrt{(\Phi^x_j-\Phi^x_{j-1})^2+(\Phi^y_j-\Phi^y_{j-1})^2}}}
\label{eq:alphadef}
\end{equation}
where $(\Phi^x_i,\Phi^y_i)$ are the coordinates of node $i$.
Periodicity of the ring (which has $N-2$ distinct nodes) is imposed by the set of algebraic equations
\begin{equation}
(\Phi_i)_t=(\Phi_{N-2+i})_t
\label{eq:bc1}
\end{equation}
where evaluation at time $t$ is indicated by the subscript outside the parentheses.
An explicit, backward in time, Euler discretization for the $N-2$
distinct nodes reads
\begin{equation}
({\Phi_i})_{t-\Delta t}=({\Phi_i})_t-\Delta t F(\Phi_i,t),\quad i=2,3,\ldots,N-1.
\label{eq:ringevolve1}
\end{equation}
Backward stepping in time is followed by a redistribution step that
slides nodes along the ring so that they are equally spaced (or,
generally, spaced in a desirable manner) in the
normalized ring arclength coordinate.
%
%
%
%
%
%
These two basic steps are also present in the (phase space $\times$ time)
arclength or potential stepping of the ring discussed below; they are
schematically summarized in Figure~5.

%
%
%
%

Figure~6
 shows snapshots of the ring as it evolves
backward in time -- in the time-stepping mode -- on the M\"{u}ller-Brown potential.
The ring quickly deviates from isopotential contours as it climbs up
the well.
The local speed is proportional to the local component of the potential gradient normal
to the ring; wide variation in nodal speeds causes the ring to
evolve unevenly, elongating along the directions of steepest ascent.
Initially equi-spaced ring nodes would, if not redistributed,
rapidly converge towards regions of high potential gradients in our
parametrization, resulting in poor resolution in other areas.
Even the redistribution of nodes, however, will not suffice to
accurately capture the ring shape as the ring perimeter quickly
grows, unless new nodes are added.


%
%
\subsubsection{Arc Length Stepping}
Integration with respect to arclength in (phase) space $\times$ time
is a well known approach for problems where some of the dependent
variables change rapidly with the independent variable (time).
Johnson et al.\cite{Johnson97} used this vector field rescaling  to
offset the concentration of flow lines in computing two-dimensional
invariant manifolds of vectorfields whose fixed points have
disparate eigenvalues.
Ring evolution by integration along the solution arc \emph{s}
%
%
is used here by transformation of the independent variable for the
system in equation~(\ref{eq:ringevolve0}).
The required transformation relation \cite{Kubicek83} is
\begin{equation}
 \left(\frac{dt}{ds}\right)_i=\left[1+\left(\frac{d\Phi^x_i}{dt}\right)^2+\left(\frac{d\Phi^y_i}{dt}\right)^2\right]^{-1/2} \equiv F_{s}(\Phi_i,t)
 \label{eq:ringevolve2} \end{equation}
with coordinates $(\Phi^x_i,\Phi^y_i)$ for node $i$.
The transformed nodal evolution equation, with solution arclength as
the independent variable, is given by
\begin{equation}
\frac{d\Phi_i}{ds}=\frac{d\Phi_i}{dt}\left(\frac{dt}{ds}\right)_i
  =F(\Phi_i,t) F_{s}(\Phi_i,t),\quad i=2,3,\ldots,N-1
\label{eq:evolve_s}
\end{equation}
where $F$ is as defined in equation~(\ref{eq:ringevolveNODE}), and the
ring boundary conditions remain periodic.

In such an arc-length stepping mode, the ring evolution for our
potential (Figure~7)
 is more robust to potential
gradient nonuniformities.
However, ring growth now does not couple to the topography
of the landscape: in Figure~7b
 it ``sags''
along the $y$-direction and there is considerable variation of
potential values along any instantaneous ring state.
%

%

\subsubsection{Potential Stepping}
Evolving in constant potential steps enables the ring to directly track
isopotential contours of the landscape.
Potential stepping is shown schematically  in one dimension
in  Figure~8
 for a potential minimum bracketed
by a sharp incline on one side and a more gradual one on the other.
A (reverse) step in the potential results in small variations in the $x$-variable
($(\Delta x)_1,(\Delta x)_2$) when the ``terrain'' is steep and in
large $x$ increments ($(\Delta x)_3,(\Delta x)_4$) when it is shallow.
A qualitatively different approach is that of Laio and Parinello
\cite{Laio02} who employed a history-dependent bias as part of free
energy surface searching that ``fills" free energy wells; using
repulsive markers actively prevents revisiting locations during
further exploration.
Irikura and Johnson \cite{Irikura00} used a combination of steps
parallel and perpendicular to the energy gradient in a version of
isopotential searching to identify chemical reaction products from a reactant
configuration.

Here we directly transform the independent variable of
the evolution equations using
the chain rule
\begin{equation}
\left(\frac{dt}{dV}\right)_i=\left[\frac{\partial V}{\partial
  \Phi^x_i}\frac{d\Phi^x_i}{dt}+\frac{\partial V}{\partial
  \Phi^y_i}\frac{d\Phi^y_i}{dt}\right]^{-1} \equiv F_{V}(\Phi_i,t)
\label{eq:energychainrule}
\end{equation}
so that, as long as the quantity above is finite (e.g. away from
critical points), the ring evolution equations now become
\begin{equation}
\frac{d\Phi_i}{dV}=\frac{d\Phi_i}{dt}\left(\frac{dt}{dV}\right)_i
  =F(\Phi_i,t) F_{V}(\Phi_i,t),\quad
  i=2,3,\ldots,N-1.
\label{eq:evolve_E}
\end{equation}
Noting that $F_{V}(\Phi_i,t)\rightarrow \infty$ in regions where the
potential is ``flat'' ($dV/dt\rightarrow 0$); we impose an upper
limit on the change in the variables $\Phi_i$ at each step of (\ref{eq:evolve_E})
when the threshold is exceeded.

%
Potential-stepping ring evolution on the M\"{u}ller-Brown potential is shown in
Figure~9:
the ring efficiently rises within the
well and successive rings are indicative of the topology of the local
landscape.
The energy well  is ``sampled'' evenly, tracking the potential
contours.
The almost linear segment  of the ring visible in the final snapshot
in Figure~9a
 is formed as the
ring approaches the (stable manifold of the) saddle point on the potential at
$(0.21,0.29)$;
no further uphill motion, normal to the ring, is possible in this
region.
When such a situation is detected, one actively intervenes and
modifies the evolution to assist the landscape search; examples of
this will be given below.
%

\subsection{Adjacent Basins}
\label{sect:basins}
The reverse integration for the example in Section
\ref{sect:ringmodes} consisted of initialization close to the bottom
of a single well, ring evolution uphill, and approach to the neighboring saddle
point.
We now discuss a reasonable strategy for transitioning between neighboring
energy wells.

Figure~10
shows the results of reverse ring integration for 3 different initial
rings, one close to the bottom of each of the wells of the M\"{u}ller-Brown
potential.
Reverse integration here maps out the basin of
attraction of each of the wells.
For each initial condition, the reverse
integration ``stalls" in the vicinity of neighboring
saddle points and ring nodes ``flow along" the stable manifold of
the saddle.
As the ring nodes approach a saddle point the component of the energy
gradient normal to the ring ($(\nabla V(\Phi_i))^\perp$) starts
becoming negligible.
To examine transitions between neighboring basins on the landscape we
can employ \emph{Global Terrain} methods \cite{Lucia04} that
exploit the inherent \emph{connectedness} of  stationary points
along valleys and ridges on the landscape.
Figure~11
indicates the basins of
attraction for each of the minima (identified using reverse
integration) along with a red curve, which connects \emph{points
that minimize the gradient norm along level curves of the
potential} (a minimum energy path).
This information is accumulated as the ring integration proceeds and suggests the
direction to follow to locate neighboring minima.
Upon detecting a local stagnation of ring evolution,
caused by the approach to a saddle, a simple strategy is to
(a) perform a local search for the saddle, through a fixed point algorithm,
(b) compute the dynamically unstable eigenvector of this saddle, and
(c) initialize a downhill search on the ``other side" along this eigenvector
{\em away from the saddle point}.
This search for the nearby minimum may be through simple forward simulation,
or (in a global terrain context) by following points that minimize
the gradient norm along level potential curves as above.
This leads to the detection of a neighboring minimum, from which a
new ring can be initialized and a further round of reverse
integration performed.
We reiterate that the procedure described so far (for purposes of
easier exposition)
is only for two-dimensional, deterministic landscapes.



%
%
%
\section{Illustrative Problems for Effective Potential Surfaces}
\label{sect:exampleprobs}

In this section we present coarse reverse integration using
effective potential stepping for three ``noisy'' problems: a
system of SDEs, a Gillespie--type stochastic simulation algorithm,
and a molecular dynamics problem (alanine dipeptide in water).
We assume that the problems we consider -- in the regime we study them
computationally -- may be effectively modelled by the
following bivariate Stochastic Differential Equation (SDE) (all the
examples studied are effectively two-dimensional)
\begin{equation}
d\bm{X}=d\left[\begin{array}{c} X_1\\X_2\end{array}\right]=\left[\begin{array}{c}
    v_1(\bm{X})\\v_2(\bm{X})\end{array}\right]dt+\left[\begin{array}{cc} D& 0\\0& D\end{array} \right]d\left[\begin{array}{c} W_{1t}\\W_{2t}\end{array}\right]
\label{eq:multidiffusionproc}
\end{equation}
where $v_1(\bm{X})$
 and $v_2(\bm{X})$
 are drift coefficients, the diffusion matrix $\bm{D}$ is proportional
 to the unit matrix $\delta_{ij}$ with $\bm{D}=D\delta_{ij}$ (a ``scalar''
 matrix),
where $D$ is a constant, and $W_{1t}$ and
$W_{2t}$ are independent Wiener processes.
We previously considered
 (Section \ref{sect:ringmodes}) a deterministic example  where
 numerical estimates for potential gradients were used to
 implement potential stepping.
In the deterministic case, the drift
 coefficients are equal to minus the gradient of a potential $V$.
For stochastic problems, such as those considered in this section,
the drift coefficients are not
 so simply related to the gradient of an effective (generalized) potential
 (see the Appendix for additional discussion of this for 1-dimensional stochastic systems).
 In general, for reverse
 integration with steps in effective potential, we require estimates of all drift
 coefficients and all
 entries in the diffusion matrix (and even their partial derivatives).
Here we discuss effective potential stepping for a system of the
form given in (\ref{eq:multidiffusionproc}) and also briefly discuss
the general case where entries of the diffusion matrix are non-zero
and dependent on $\bm{X}$.

We assume equation~(\ref{eq:multidiffusionproc}) exists but
is unavailable in closed form; estimates are therefore obtained
 by observing the process $\bm{X}$
and using
$v_1(\bm{X})\equiv \lim_{\Delta
   t\rightarrow 0}\langle\left[\Delta X_1\right]\rangle/\Delta t,
v_2(\bm{X})\equiv \lim_{\Delta
   t\rightarrow 0}\langle\left[\Delta X_2\right]\rangle/\Delta t$,
and  $2D\equiv \lim_{\Delta
   t\rightarrow 0}\langle\left[\Delta X_1\right]^2\rangle/\Delta t=
 \lim_{\Delta
   t\rightarrow 0}\langle\left[\Delta X_2\right]^2\rangle/\Delta t$.
Here $\Delta X_i=X_i(t+\Delta t)-X_i(t)$ and, by the form of
equation~(\ref{eq:multidiffusionproc}), $\lim_{\Delta t\rightarrow
0}\langle\Delta X_1\Delta X_2\rangle/\Delta t=0$.
%
%

These formulas, especially the ones for the drifts, suggest the
construction of a useful coarse ``pseudo-dynamical" evolution for
our ring –- a coarse evolution that follows the potential of mean
force.
The simplest version of these pseudo-dynamics evolves each point
 on the ring based on the local estimated drift -- for the constant
 ``scalar" diffusion mentioned above this evolution follows the potential of mean force (PMF),
  and it becomes a true dynamical evolution at the
deterministic limit.

For a black box code implementing
equation~(\ref{eq:multidiffusionproc}) this involves initializing
at $\bm{X}$, running an ensemble of realizations of the dynamics
for a short time $\delta t$, estimating the local drift components
of the SDE using the above formulas,  performing a (forward or
backward) projective step $\Delta t$ in time ($\Delta
X_i=v_i(\bm{X})\Delta t$), and repeating the process.

We will argue that this accelerated pseudo-dynamical evolution
(which, we emphasize, does {\it not} correspond to realizations of the SDE itself)
can assist in the exploration of effective potential surfaces.
The easiest approach would be to use reverse time-stepping, or reverse arclength
 stepping in these pseudo-dynamics,
and then (using formulae that will be discussed below and in the
Appendix) finding the effective potential corresponding to each
node visited.
It is also possible, as we will see, to directly make ``upward"
steps in the effective potential; indeed, for the constant
diffusion coefficient case we are studying, a proportionality
exists between backward steps in time (for the pseudo-dynamics
based on the drifts) and upward steps in the effective potential.

In the SDE case (Section \ref{sect:sdesystem}) we only allow
ourselves to observe sample paths generated by
short bursts of the SDE solver; the SDE solver itself is treated
as a black box (similarly for the Gillespie and MD simulators).
A simple approach to estimating
effective potential gradients (and eventually
free energy gradients) is to
perform sets of $M$-replica bursts of {\em inner} (SDE, Gillespie, MD) simulation
initialized at each of the $N$ ring nodes.
For short replica simulation bursts (with $n$ time steps), we can assume a local
first order in time model\cite{Drape81} for the {\em mean} $\bm{\overline{x}}$
(an $n\times2$ matrix, with entries averaged using multiple replicas,
rows corresponding to time abscissas, and columns corresponding to each coarse variable)

\begin{equation}
\bm{\overline{x}}=\tilde{\bm{t}}\bm{C}+{\bm \epsilon}
\label{eq:firstordermodel}
\end{equation}
where ${\tilde{\bm t}}=\left[\bm{1}\  {\bm t}\right]$ is a $n\times2$ matrix, $\bm{1}$ is a
vector of  $n$ ones, $\bm{t}$ is a vector of time abscissas, ${\bm
  \epsilon}$ is the $n\times2$ matrix of model errors, and ${\bm C}$ is the $2\times2$
matrix of parameters computed (for each node) using least squares estimation.
The (pseudo-time) derivative information (in the matrix $\bm C$) is
required, along with approximations of the tangent vectors at each
node (ring geometry) to update the ring node positions in a reverse
integration step; diffusion coefficients are also required, as
discussed further below, to compute the relation between a reverse
integration step size in pseudo-time and the corresponding change in the
effective potential.
In the remainder of the paper reverse ring time-stepping
is always meant in terms of the drift-based pseudo-dynamics (it
only becomes true time-stepping at the deterministic limit).

This derivative information may also be used to confirm the existence of an
 effective potential. For the case of two effective coarse-dimensions,
we locally compare, computing on a stencil of
  points, the $X_2$-variation of $dX_1/dt$ with the $X_1$-variation of
 $dX_2/dt$  (testing for equality of mixed
  partial derivatives of the effective potential). Alternatively, we may use a
 locally affine model for the drift coefficients of the following form
\begin{equation}
\left[\begin{array}{c}
    v_1(\bm{X})\\v_2(\bm{X})\end{array}\right]=\bm{A}\bm{X}+\bm{B}
\label{eq:driftaffine}
\end{equation}
with $\bm{A}\in\mathbb{R}^{2\times2}$, $\bm{B}\in\mathbb{R}^2$, and
 employ maximum likelihood estimation techniques to compute $\bm{A}$ and
 $\bm{B}$ (an effective potential exists provided $A_{12}=A_{21}$).
In this context, recently developed maximum
likelihood\cite{AitSahalia02} or
Bayesian\cite{Hummer05} estimation approaches are particularly
promising, allowing for simultaneous estimation of {\em both} the drift and diffusion coefficients.
%
These approaches assume that the
data are generated by a (multivariate) parametric
diffusion; they employ a closed-form approximation to the transition
density for this diffusion.
For the case of a one-dimensional diffusion process $\tilde{X}$
\begin{equation}
d\tilde{X}=\mu(\tilde{X};\theta)dt+\sigma(\tilde{X};\theta)dW_t
\label{eq:diffusionproc}
\end{equation}
where $W_t$ is the Wiener process, $\theta$ is a parameter vector, $\mu$
is the drift coefficient, and $\sigma$ is the diffusion coefficient,
the corresponding log
likelihood function $l_n(\theta)$  is defined as

\begin{equation}
l_n(\theta)=\sum_{i=1}^n ln\left[p_{\tilde{X}}(\Delta,\tilde{X}_{i\Delta}\mid \tilde{X}_{(i-1)\Delta};\theta)\right]
\label{eq:loglikelihood}
\end{equation}
where $n$ is the number of time abscissas, $\tilde{X}_{i\Delta}$ is
the $i^{th}$  sample, and $\Delta$ is the time step
between observations in the time series.
The derivation of a closed-form
expression for the transition density $p_{\tilde{X}}$ (and thereby the log likelihood function) allows
for maximization of $l_n$ with respect to the parameter vector
$\theta$ providing ``optimal'' estimates for the drift and
diffusion coefficients associated with the time series.
For higher-dimensional problems (such as the two-dimensional ones considered here)
 see Ref.~48.
%
%

If the system in (\ref{eq:multidiffusionproc}), with ``scalar'' diffusion
matrix, has drift coefficients that satisfy the following
 {\em potential condition}
\begin{equation}
\frac{\partial v_1(\bm{X})}{\partial X_2}=\frac{\partial
  v_2(\bm{X})}{\partial X_1}
\label{eq:potentialconds}
\end{equation}
it follows that the probability current vanishes at equilibrium, the drift
 coefficients (the time-derivatives in our ring pseudo-dynamics) satisfy
 \begin{equation}
 v_i=-D\frac{\partial(\beta
 E^{\mathrm{eff}})}{\partial X_i},
\label{eq:driftrelation}
\end{equation}
 and the difference in effective
 generalized potential (free energy) between a reference state $(X_1^0,X_2^0)$ and
 the state $(X_1,X_2)$ may be directly
  computed from the following line integral\cite{Risken96}
\begin{equation}
\beta
\Delta E^{\mathrm{eff}}=-D^{-1}\left(\int_{X_1^0}^{X_1}v_1(X_1',X_2^0)dX_1'+\int_{X_2^0}^{X_2}v_2(X_1,X_2')dX_2'\right).
\label{eq:lineint1}
\end{equation}
The analogy with the deterministic case
(eqs.(\ref{eq:energychainrule}) and (\ref{eq:evolve_E})) carries
through: the estimated drifts are proportional (via the constant
$D$) to the effective potential gradients, and evolution following
the drifts directly corresponds (modulo the proportionality
constant) to evolution in the effective potential (PMF).
Estimates of the local effective diffusion
coefficients are typically necessary for exploration of the effective
potential surface.
We note that for a diagonal
diffusion tensor with identical entries, (\ref{eq:multidiffusionproc}),
the size of the step $\beta\Delta E^{\mathrm{eff}}$ is scaled (in (\ref{eq:lineint1}))
by the diffusion constant $D$.
It follows that estimation of only the drift coefficients
$v_1(\bm{X})$ and $v_2(\bm{X})$ allows us to perform reverse
integration in our coarse dynamics (associated with the potential
of mean force).
A backward in time step $\Delta t$, leading to the state change
$\Delta X_i=v_i(\bm{X})\Delta t$, is, in effect, an ``upward" step
in the effective potential with the (unknown) scaled stepsize
$D\beta\Delta E^{\mathrm{eff}}$.
This approach is analogous to (and, in the appropriate limit will
approximate) the deterministic potential stepping previously
described (Section \ref{sect:ringmodes}).
Here, for a stochastic
problem, we need to additionally estimate diffusion coefficients to
compute the potential change associated with each ring step uphill
and, thereby, the effective free energy change associated with each
ring.
%
%

%
For the general diffusion matrix $\bm{D}(\bm{X})$, with
all entries possibly non-zero and dependent on $\bm{X}$, we would compute the
following partial derivatives of the effective potential
\begin{eqnarray}
\frac{\partial(\beta E^{\mathrm{eff}})}{\partial X_1}
\equiv A_1
=
\left(\bm{D}(\bm{X})\right)^{-1}_{11}
\left(\frac{\partial D_{11}}{\partial X_1}+\frac{\partial
  D_{12}}{\partial X_2}-v_1\right)\nonumber\\
+
\left(\bm{D}(\bm{X})\right)^{-1}_{12}
\left(\frac{\partial D_{21}}{\partial X_1}+\frac{\partial
  D_{22}}{\partial X_2}-v_2\right)\\
\frac{\partial(\beta E^{\mathrm{eff}})}{\partial X_2}
\equiv A_2
=
\left(\bm{D}(\bm{X})\right)^{-1}_{21}
\left(\frac{\partial D_{11}}{\partial X_1}+\frac{\partial
  D_{12}}{\partial X_2}-v_1\right)\nonumber\\
+
\left(\bm{D}(\bm{X})\right)^{-1}_{22}
\left(\frac{\partial D_{21}}{\partial X_1}+\frac{\partial
  D_{22}}{\partial X_2}-v_2\right)
\end{eqnarray}
and test whether the following potential condition is satisfied\cite{Risken96}
\begin{equation}
\frac{\partial A_1(\bm{X})}{\partial X_2}=\frac{\partial
  A_2(\bm{X})}{\partial X_1}.
\label{eq:potentialconds2}
\end{equation}
If these potential conditions are satisfied then the effective
 generalized potential (free energy) may again be directly calculated from the following line
 integral\cite{Risken96}
\begin{equation}
\beta E^{\mathrm{eff}}(X_1,X_2)=\beta
E^{\mathrm{eff}}(X_1^0,X_2^0)+\int_{X_1^0}^{X_1}A_1(X_1',X_2^0)dX_1'+\int_{X_2^0}^{X_2}A_2(X_1,X_2')dX_2'.
\label{eq:lineint2}
\end{equation}
We do not consider the case when equation
(\ref{eq:potentialconds2}) does not hold; we refer the reader to
Ref.~49.
%
%

%
%
%
In the same spirit with reverse ring stepping in potential (Section
\ref{sect:ringmodes}), reverse ring stepping in effective potential
may also be accomplished, subject to the stated assumptions, using
the inner integrator as a black-box: we run multiple replicas for
particular initial conditions (the positions of nodes in the ring),
observe (inner) forward time evolution, and, for a ``scalar''
diffusion matrix, use the estimated drifts and (\ref{eq:lineint1})
to approximate changes in the effective potential numerically.
We note that for a constant and isotropic diffusion tensor if we
estimate {\em only} the drift coefficients we can still perform
reverse ring stepping in the correct uphill direction and follow
isopotential surfaces but the actual step size (and thus the
actual value of the potential on the isopotential surfaces) will
be unknown. As reverse ring integration proceeds, we store all
calculated effective gradient values at each set of coarse
variable values, thereby building a database. Smoothed gradient
estimates may be obtained for each ring node by using a weighted
gradient average that includes estimates at nearby coarse variable
values in the database; we use kernel smoothing\cite{Wand95} to
select appropriate weights. For the more general case of
state-dependent diffusion the drift dynamics do not simply
correspond to dynamics in the effective potential (see the
Appendix for corrections to $d\Phi_i/dt$ required to retain the
analogy to the deterministic equations (\ref{eq:energychainrule})
and (\ref{eq:evolve_E})).
One could still employ the uncorrected drift dynamics as an
 \emph{ad hoc} search tool (especially for problems close to ``scalar"
diffusion matrices) and post-compute the effective potential
values the ring visits.
In this case, however, the time-parametrization of the effective
potential evolution will not be meaningful, and will even
dramatically fail in the neighborhood of drift steady states that
do not correspond to critical points in the effective potential
(and vice versa).

%
\subsection{A Stochastic Differential Equation Example}
\label{sect:sdesystem} In this section we consider ring evolution
in potential-stepping mode for a system of stochastic differential
equations (SDEs).
Reverse ring
integration is performed at the \emph{outer} level.
The \emph{inner} routine here is a forward-in-time SDE
(Euler-Maruyama) integrator based on which we generate the nodal
gradient estimates required by the outer ring integrator.
The SDE system is given by
\begin{equation}dx(t)=D_x F_x(t)\ dt+\sqrt{2D_x}\ dW_{1t}
\label{eq:SDE1}
\end{equation}
\begin{equation}dy(t)=D_y F_y(t)\ dt+\sqrt{2D_y}\ dW_{2t}
\label{eq:SDE2}\vspace{0.25cm}
\end{equation}
where $D_x=D_y=D=1, F_x=-\frac{\partial V}{\partial
  x},F_y=-\frac{\partial V}{\partial y}$, and the function $V(x,y)$ is given by
\begin{equation}
V(x,y)=10(x^2-1)^2+2x+\tfrac{1}{2}(y-x)^2.
\label{eq:SDE_V(xy)}
\end{equation}
%
%
The discretized (using
the Euler-Maruyama scheme) system of equations is as follows
\begin{equation}
x(t+\Delta t)=x(t)+{D}F_x\Delta t+\sqrt{2D\Delta
  t}\ \mathcal{N}(0,1)\label{eq:SDEdisc1}
\end{equation}
\begin{equation}
y(t+\Delta t)=y(t)+{D}F_y\Delta t+\sqrt{2D\Delta
    t}\ \mathcal{N}(0,1)\label{eq:SDEdisc2}
\end{equation}
where $\mathcal{N}(0,1)$ is a normal random number with mean 0 and
unit variance.
We initialize the ring on an isopotential contour about the
minimum at $(-1.024,-1.024)$.
The ratio of eigenvalues at this fixed
point is approximately 90 -- the well is sharply elongated in the
$y$-direction.
To cope with this sharp elongation, we adaptively adjust the
distribution of ring nodes so that they remain concentrated in
regions where the ring curvature is largest (we did not adaptively
change the number of nodes here).

The results of ring evolution (following drifts only) in a single
well for the SDE problem are shown in the left panel of
Figure~12
(contour lines are shown for $V(x,y)$).
Here we plot ring nodal positions at every reverse integration
step in drift potential. Here, the selected diffusion coefficients
($D_x=D_y=D=1$) and the functional form of the drift and diffusion
terms in eqs.(\ref{eq:SDE1}) and (\ref{eq:SDE2}) necessarily imply
that $\beta=1$ -- the effective potential is essentially the same
as the drift potential (see also the Appendix). Reverse
integration eventually stalls in the vicinity of the saddle point
at $(0,0)$. In Figure~12
 (right panel) we show the
estimated effective potential associated with ring nodes
superimposed on 3D contour lines for $V(x,y)$. Estimates of both
local drift and diffusion coefficients are used to compute
effective potential differences (using
equation~(\ref{eq:lineint1}))
 for successive rounds of reverse integration
 (as generated by potential stepping, shown left). The effective
 potential is computed relative to that of the initial condition
 (a ring on an isopotential contour).

As the local curvature of the landscape changes, the duration of our
short inner computation bursts (the time interval over which we
collect data to estimate derivatives) should be adaptively modified
for computational accuracy.
%
%

%
%
%
%
%

\subsection{A Gillespie--type SSA inner simulator example}
\label{sect:gillespiesystem}
The stochastic description of a spatially homogeneous set of
chemical reactions, which treats the collisions of species in the
system as essentially random events, is based on the chemical
master equation\cite{McQua67}.
The Gillespie Stochastic Simulation Algorithm (SSA) is a Monte
Carlo procedure used to simulate a stochastic formulation of
chemical reaction dynamics that accounts for inherent system
fluctuations and correlations -- this procedure numerically
simulates the stochastic process described by the spatially
homogeneous master equation \cite{Gillespie77}.
At each step in the simulation a reaction event is
selected (based on the reaction probabilities), the species numbers
updated (according to the
stoichiometry of the reactions) and the time to the next
reaction event computed.
The reaction probabilities used in the algorithm are determined by
the species concentrations and reaction rate constants as
described in Ref.~52.
%
The inner stochastic simulation routine we use here happens to
employ an explicit tau-leaping scheme that takes larger time steps
to encompass more reaction events while still ensuring that none
of the propensity (reaction probability) functions in the
algorithm changes significantly\cite{Gillespie01}.
The reaction events we simulate are chosen to implement a mechanism
which, at the limit of infinitely many particles, would be described
by the deterministic gradient system with potential $V(x,y)$ defined in
 equation~(\ref{eq:SDE_V(xy)}).

Consider the following deterministic rate
equations
\begin{align}
\frac{dx}{dt}&=-{k_1}x+{k_2}{x^2}-{k_3}{x^3}+k_4-k_5x+k_6y
\label{eq:ODE1}\\
\frac{dy}{dt}&={k_5}x-{k_6}y+k_7.
\label{eq:ODE2}
\end{align}
This set of deterministic (coarse) rate equations may be written, for
this problem, in
the form of the following gradient system

\begin{equation}
\frac{d\bm x}{dt}=-\nabla V^*(x,y)
\label{eq:GradientSystem}
\end{equation}
where ${\bm x}$ may be interpreted, here, as a vector of
chemical species concentrations, and the potential energy function
$V^*(x,y)$ is given by
\begin{equation}
V^*(x,y)=\frac{(k_1+k_5)}{2}\ x^2-\frac{k_2}{3}\ x^3+\frac{k_3}{4}\
x^4-k_4x-k_5xy+\frac{k_6}{2}\ y^2-k_7y+k_8
\label{eq:Eshifted}
\end{equation}
with
\begin{equation}
k_5=k_6.
\end{equation}
Values for the rate constants are selected by requiring
$V^*(x,y)=V(x-5,y-20)$ (i.e. $V^*(x,y)$ is selected as a shifted
version of the $V(x,y)$ from the previous example, with its fixed
points in the positive $xy$ quadrant, in an attempt to enforce
positivity of the reaction probabilities required by the Gillespie
algorithm).
The rate constant values chosen are $k_1=2960,
k_2=600,k_3=40,k_4=4783,k_5=k_6=1,k_7=15$. This models the
following, hypothetical, set of elementary reactions
\begin{align}
X &\stackrel{k_1}{\rightarrow}\overline{T} \label{eq:Grxn1}\\
2X+\overline{U}&\overset{k_2}{\underset{k_3}{\rightleftharpoons}}3X \label{eq:Grxn2}\\
\overline{V} &\stackrel{k_4}{\rightarrow}X \label{eq:Grxn4}\\
X &\overset{k_5}{\underset{k_6}{\rightleftharpoons}}Y \label{eq:Grxn5}\\
\overline{W}&\stackrel{k_7}{\rightarrow}Y \label{eq:Grxn7}
\end{align}
where species $X$(resp.\ $Y$) has concentration $x$(resp.\ $y$),
 the species $T, U, V, \mbox{and } W$ are
assumed to have unchanging concentration 1, and the reactions
 in eqs.(\ref{eq:Grxn4}) and
(\ref{eq:Grxn7}) follow zeroth order kinetics.

For the number of particles used in this Gillespie simulation, the
drift coefficients estimated from the simulation practically
coincide with the right-hand-side of the {\it deterministic} rate
equations, which happen to be embody the gradient of the {\it
deterministic} potential $V(x,y)$ (equation~(\ref{eq:SDE_V(xy)})).
The results of reverse ring integration up this deterministic
potential, with drifts estimated from our Gillespie simulation are
shown in Figure~13.
%
The left panel shows nodal
evolution over 100 rounds of reverse integration. In the right panel
we superimpose the nodal evolution (with estimated potential
indicated by color) on contours of the potential $V(x,y)$ (defined
in equation~(\ref{eq:SDE_V(xy)}))
 for the deterministic gradient system in the form of
 (\ref{eq:gradientsys}).
Since we are using an explicit tau-leaping Gillespie scheme, we do
not have accurate estimates of the diffusion coefficients of the
underlying chemical Fokker-Planck equation \cite{Cao04}.
For this problem these entries in the diffusion matrix cannot be
well approximated as state-independent, and a more involved process that includes their
estimation is required in order to construct the true effective
potential.
%

%
%
%

\subsection{Alanine dipeptide in Water}
\label{sect:aladip} In this section we study the coarse effective
potential landscape of alanine dipeptide (i.e. $N$-acetyl alanine
$N'$-methyl amide) dissolved in water using coarse reverse
(effective potential-stepping) integration.
This system is a basic fragment of protein backbones with two main torsion angle degrees of freedom
$\phi$ ($C-N-C_{\alpha}-C$) and $\psi$ ($N-C_{\alpha}-C-N$), and with polar groups that
interact strongly with each other and with the solvent.
Extensive theoretical and experimental investigation of the
alanine dipeptide has suggested good coarse observables (dihedral
angles) for this system \cite{Rossky79,Brooks93,Chipot98}.

%
Figure~14
 shows the effective free energy
landscape as a function of the dihedral angles $\phi$  and $\psi$
of the alanine dipeptide.
The structures of the alanine dipeptide in the $\alpha$-helical
($\psi~=-0.3$) and extended ($\psi=\pi$) states (corresponding to
minima on the landscape) and at the transition state between them
are also shown.
We will use
reverse integration on the effective potential energy landscape parametrized by these coarse coordinates.
The coarse reverse integration is ``wrapped around'' a conventional
forward-in-time molecular dynamics (MD) simulator.
It provides protocols for where (i.e. at what starting values of the coarse variables)
to execute short bursts of MD, so as to map the main
features of the effective potential surface (minima and connecting
saddle points).
These short bursts of appropriately initialized MD simulations
provide (via estimation of the coefficients in equation
(\ref{eq:multidiffusionproc})) the deterministic and stochastic
components of the alanine dipeptide coarse dynamics parametrized
by the selected coarse variables. The current work assumes a
diffusion matrix (equation (\ref{eq:multidiffusionproc})) that is
diagonal with identical constant entries.
Our MD simulations of the alanine dipeptide in explicit water are
performed using AMBER 6.0 and the parm94 force field.
The system is simulated at constant
volume corresponding to 1 bar pressure, and the temperature is
maintained at 300K by weak coupling to a Berendsen thermostat.
All simulations
use a time step of 0.001 ps.
The ``true" effective potential here is the one obtained from the
stationary probability distribution as approximated by a long MD
simulation (24 ns).

A preparatory ``lifting'' step is required at each reverse integration
step for each ring node.
Each coarse initial condition is lifted
to many microscopic copies conditioned on the coarse variables $\phi$ and $\psi$.
This
step is not unique, since many distributions may be constructed having
the same values of the coarse variables.
Here we lift by performing a
short MD run with an added potential $V^{\mathrm{constr}}$ that biases
(as in umbrella sampling) the coarse variables towards their target values
($\psi^{\mathrm{targ}}$,$\phi^{\mathrm{targ}}$),

\begin{equation}
V^{\mathrm{constr}}=k_{\psi}(\psi-\psi^{\mathrm{targ}})^2/2+k_{\phi}(\phi-\phi^{\mathrm{targ}})^2/2
\label{eq:constrpot}
\end{equation}
with $k_{\psi}=k_{\phi}=100$ $\mathrm{kcal\ mol^{-1} rad^{-1}}$.
The short lifting phase provides sufficient time for the fast variables
to equilibrate following changes in the coarse variables.
Following initialization we run and monitor the detailed MD simulations over
short times (0.5ps) and estimate, for each node in the coarse variables, the local drifts
over multiple replicas.
%
%
Each coarse backward Euler step
of the ring evolution provides new coarse variable values at which to
initialize short bursts of the MD simulator.
Each step in the reverse integration procedure consists of lifting
from coarse variables (the coordinates of the ring nodes) to an
ensemble of consistent microscopic configurations, execution of
multiple short MD runs from such configurations, restriction to
coarse variables, estimation of coarse drifts and diffusivities,
and reverse Euler stepping of the ring in the chosen evolution
mode.


Figure~15
 (left panel) shows ring nodes for 30 steps of reverse ring integration
(using $N$=12 nodes) initialized around the extended structure minimum.
Successive rings
evolve up the well and  are representative  of the
well topology.
Reverse integration stalls, as expected, at the saddle points neighboring the
extended structure minimum and identifies
candidate saddle points in these regions.
We note that, in the current context of (assumed) constant diffusion
coefficients we can think of these saddles as steady states of the set
of deterministic ODEs, coinciding with the drift terms of the
effective Fokker-Planck.
Then the ``dynamically unstable'' directions in a saddle (the
downhill ones) are characterized by {\em positive} eigenvalues of
the {\em Jacobian} of the drift equations; yet since these
equations are proportional to the negative of the gradient of a potential,
positive eigenvalues of the dynamical Jacobian correspond to
negative eigenvalues of the Hessian.
The eigenvectors associated with the unstable (for our PMF-related
coarse dynamics) eigenvalue at these candidate saddles are also
indicated in Figure~15
 and suggest the
directions to dynamically follow to locate neighboring minima.
We perturbed in the direction of the unstable
eigenvector (associated with positive eigenvalue) away from one of the
candidate saddle points and initialized (using a constrained potential,
as before) multiple MD runs from this location.
In Figure~15
 (right panel) we plot the observed
evolution from these initial conditions down into the basin of the adjacent
$\alpha$-helical minimum.

%

%
In Figure~16
 we show reverse ring evolution
initialized close to both $\alpha$-helical and extended minima.
Clearly reverse ring evolution in this $\alpha$-helical minimum
well takes larger steps in $\phi$, in which direction the
effective potential is shallowest.
We repeat that the reverse integration steps correspond to
constant steps in free energy only if the effective diffusion
tensor is diagonal and constant in both directions.
The ring evolution shown in Figure~16
 appears to accurately track equal free
energy contours suggesting that these assumptions (on the form of
the diffusion tensor) are a suitable approximation here.

\section{Summary and Conclusions}

We have presented a coarse-grained computational approach (coarse reverse integration) for
exploration of low-dimensional effective landscapes.
In our two-coarse-dimensional examples an (outer) integration scheme evolves a
ring of replica simulations backwards by exploiting short bursts of a
conventional forward-in-time (inner) simulator.
The results of small periods of forward inner simulation  are
processed to enable large steps backward in time (pseudo-time in the
stochastic case), in phase space
$\times$ time, or in potential in the outer integration.
We first illustrated these different modes of reverse integration for smooth, deterministic landscapes.
We extended the most promising approach for an illustrative
deterministic problem, isopotential stepping, to relatively simple
noisy (or effectively noisy) systems where closed-form evolution
equations are not available.
Simple estimation techniques were applied here to the results of appropriately
initialized short bursts of forward simulation used locally to extract stochastic models
with constant diffusion coefficients.
Reverse integration in a single well and the approach
to/detection of neighboring coarse saddles was demonstrated.
A brief discussion of Global Terrain approaches for exploring potential surfaces
was included, along with a short demonstration of linking our approach
to them.

We have presented here ring exploration using an effective potential,
using only estimation of the drift coefficients of our
effective coarse model equations.
Estimation of the
diffusion coefficients (and their derivatives) is additionally required to
quantitatively trace the effective potential surface.
More sophisticated estimation techniques
\cite{AitSahalia02b,Hummer05} allow for reliable estimation of both
the stochastic and deterministic components of the coarse model
equations.
This permits a quantitative reconstruction of the effective free
energy surface (and thereby the equilibrium density) using our
reverse integration approach.
The latter reconstruction is possible provided that the potential
conditions discussed in Section \ref{sect:exampleprobs} hold;
testing this hypothesis should become an integral part of the
algorithm.

In studies of high-dimensional systems, a central question is the
appropriate choice of coarse variables used in the reverse
integration.
For high-dimensional systems, such as those arising in molecular
simulations, the dynamics can typically be monitored only along a
few chosen ``coarse" coordinates.
Formally, an exact evolution equation can be derived for these
coordinates with the help of the projection-operator formalism
\cite{Zwanz01}, but that equation will be non-Markovian even if
the time evolution in the full space is Markovian.
To minimize the resulting memory effects, one can attempt to
identify good (i.e., nearly Markovian) coordinates {\em a priori},
e.g., based on the extensive experience with the problem (as, say,
in hydrodynamics) or by data analysis \cite{Nadle06,Best05}.
Alternatively, one can monitor the dynamics in a large space of
trial coordinates and select a suitable low-dimensional space on
the fly (e.g. from Principal Component Analysis\cite{Garcia92}).
In general problems, where good coordinates are not immediately
obvious, careful testing of the Markovian character of the
projected dynamics on the time scale of the coarse forward or
reverse integration will be an important component of the
computation\cite{Best05,Hummer07}.

For the alanine dipeptide in water many-degree-of-freedom example,
we assumed that the effective dynamics could be described in terms
of a few coarse variables known from previous experience with the
problem: the two dihedral angles.
%
%
We are also exploring the use of diffusion map techniques
\cite{Coifm05} for data-based detection of such coarse
observables, in effect trying to reconstruct
Fig.\ref{fig:aladip_ringall} without previous knowledge of the
dihedral angle coarse variables.
An example of mining large data sets from protein folding
simulations to detect good coarse variables using a scaled Isomap
(ScIMAP) approach can be found in Ref.~64; linking coarse
variables with reverse integration for this example is discussed
further in an upcoming publication\cite{Das07}.
All the work in this paper was in two coarse dimensions.
In the context of invariant manifold computations for dynamical systems (which
provided the motivation for this work) more sophisticated algorithms exist for the
computer-assisted exploration of  higher-dimensional
manifolds (as high as 6-dimensional) \cite{Hende02,Hende05}.
It should be possible -- and interesting ! -- to use these
manifold parametrization and approximation techniques in
combination with the approach presented here, to test the ``coarse
dimensionality" of effective free energy surfaces one can usefully
explore.

{\bf Acknowledgements.} This work was partially supported by DARPA
and NSF (TAF and IGK) and by the Intramural Research Program of
the NIDDK, NIH (GH).

\clearpage

\renewcommand{\theequation}{A-\arabic{equation}} 
\setcounter{equation}{0}  

\section*{Appendix}
\textbf{Stationary Probability Distribution and Effective Free
Energy}

We discuss here the effective potential (effective free energy
$E^{\mathrm{eff}}(\psi)$) we attempt to compute through reverse
integration and its relation to the form of the stationary
probability distribution $P_{st}(\psi)$ for a 1-dimensional Fokker
Planck equation (FPE).
%

In 1-D we write the FPE (with drift
\begin{equation}
v(\psi_0)=\frac{\partial<\psi(t;\psi_0)>}{\partial t}
\end{equation}
 and diffusion coefficient
\begin{equation}
D(\psi_0)=\frac{1}{2}\frac{\partial\sigma^{2}(t;\psi_0)}{\partial
t},
\end{equation}
where $\psi(t;\psi_0)$ is a sample path of
duration $t$ initialized at $\psi_0$ when $t=0$ and where
$\sigma^{2}(\psi_0,t)$ is the variance of $\psi(t;\psi_0)$) as
follows:

\begin{equation}\label{FPE}
    \frac{\partial P(\psi,t)}{\partial t}=
    \left[-\frac{\partial}{\partial\psi}v(\psi)+
    \frac{\partial^{2}}{\partial\psi^{2}}D(\psi)\right]P(\psi,t)=
    -\frac{\partial S(\psi,t)}{\partial\psi}
\end{equation}
where the probability current $S(\psi,t)$ is given by
\begin{equation}\label{ProbCurr}
S(\psi,t)=v(\psi)P(\psi,t)-(\partial/\partial\psi)D(\psi)P(\psi,t).
\end{equation}
In 1-D, the stationary probability distribution corresponds to a
constant probability current \cite{Risken96}; for natural boundary
conditions this constant is zero and stationary solutions of the
FPE satisfy

\begin{equation}\label{ProbCurrConst}
v(\psi)P_{st}(\psi)-(\partial/\partial\psi)D(\psi)P_{st}(\psi)=0
\end{equation}
which is readily solved for (the logarithm of) $P_{st}$

\begin{equation}\label{ProbStat}
    ln P_{st}(\psi)=-ln D(\psi)
    +\int^{\psi}\frac{v(\psi')}{D(\psi')}d\psi'+const.
\end{equation}
The connection between the stationary probability distribution and
the effective potential (effective free energy) for systems with a
characteristic temperature/energy scale (given by the parameter
$\beta^{-1}=k_{B}T$), is provided by the ansatz
$P_{st}(\psi)\propto e^{-\beta E^{\mathrm{eff}}(\psi)}$.
Substitution of the ansatz into eq.(\ref{ProbStat}) gives

\begin{equation}\label{PotEff}
    \beta E^{\mathrm{eff}}(\psi)=ln D(\psi)-\int^{\psi}
    \frac{v(\psi')}{D(\psi')}d\psi'+const.'
\end{equation}

In Section 3, after the fitting of model SDEs, we discussed the
use of local estimates of the drift and diffusion coefficients in
taking steps in some form of the effective potential for 3 example
systems; we consider the basis of this approach here in 1-D.
For both the SDE and Gillespie problems of Section 3 reverse
ring stepping results were compared to particular
deterministic potentials $V(\psi)$ (eq.(\ref{eq:SDE_V(xy)})).
For the alanine dipeptide problem the results of reverse ring
stepping were compared to an effective potential derived from the
stationary probability distribution of the system (with the
additional assumption of state-independent diffusion
coefficients).

If, alternatively, we start from the Langevin equation
\begin{equation}\label{Langevin}
    \ddot{\psi}=-\gamma(\psi)\dot{\psi}+f_{0}(\psi)+\Gamma(t).
\end{equation}
where $\gamma(\psi)$ is the friction coefficient, $f_{0}(\psi)$ is
a deterministic force (minus the gradient of a deterministic
potential function $V(\psi)$), $\gamma(\psi)\dot{\psi}$ is a drag
force, and $\Gamma(t)$ is the stochastic force, and take the high
friction (overdamped) limit we obtain

\begin{equation}\label{Langevin}
    \dot{\psi}=\frac{f_{0}(\psi)}{\gamma(\psi)}+\frac{\Gamma(t)}{\gamma(\psi)}.
\end{equation}
The fluctuation-dissipation relation connects (correlations of)
the stochastic force to the drag force as follows

\begin{equation}\label{FluctD}
    \langle\Gamma(t)\Gamma(t+\tau)\rangle=\frac{2\gamma(\psi)\delta(t-\tau)}{\beta}
\end{equation}
for a system at ``temperature" $T$ (energy scale $k_{B}T=\beta^{-1}$).
Using Ito calculus we interpret eq.(\ref{Langevin}) as
\begin{equation}\label{SDE0}
    d\psi=\frac{f_{0}(\psi)}{\gamma(\psi)}dt+\sqrt{\frac{2k_{B}T}{\gamma(\psi)}}dW_{t}
\end{equation}
with

\begin{equation}\label{driftIto}
    v(\psi)\equiv\frac{f_{0}(\psi)}{\gamma(\psi)}
    =-\frac{1}{\gamma(\psi)}\frac{dV(\psi)}{d\psi}
\end{equation}

\begin{equation}\label{diffusionIto}
    D(\psi)\equiv\frac{k_B T}{\gamma(\psi)}=\frac{1}{\beta\gamma(\psi)}.
\end{equation}
This establishes a correspondence of the Langevin equation with
the FPE in eq.(\ref{FPE}).

For the case of additive noise, where $D(\psi)=D=const$ (implying
(by eq.(\ref{diffusionIto})) that $\gamma(\psi)=\gamma=const$), we
find (by differentiation of eq.(\ref{PotEff}) w.r.t. $\psi$) that
the drift coefficient $v(\psi)$ is simply related to the effective
potential $E^{\mathrm{eff}}$ as follows

\begin{equation}\label{StateInd}
    v(\psi)\left(=\frac{d<\psi>}{dt}\right)=
    -\frac{1}{\gamma}\frac{dE^{\mathrm{eff}}(\psi)}{d\psi}
    =-D\frac{d(\beta E^{\mathrm{eff}}(\psi))}{d\psi}.
\end{equation}
In this case, pseudo-dynamical reverse integration following drifts (as performed
for the model SDE problem) coincides with stepping in effective
potential (appropriately scaled with the constant diffusion
coefficient). Using eq.(\ref{driftIto}) in eq.(\ref{StateInd}) we
find

\begin{equation}\label{StateInd}
    \frac{d(\beta V(\psi))}{d\psi}=\frac{d(\beta E^{\mathrm{eff}}(\psi))}{d\psi}\
    =-\frac{1}{D}\frac{d<\psi>}{dt}.
\end{equation}

For the case of multiplicative (state-dependent) noise the drift
coefficient $v(\psi)$ is not directly related to the gradient of
the effective potential $E^{\mathrm{eff}}$ extracted from the
equilibrium density; instead, it satisfies

\begin{equation}\label{StateDep}
    v(\psi)\left(=\frac{d<\psi>}{dt}\right)=
    -\frac{1}{\gamma(\psi)}\frac{d\hat{E}^{\mathrm{eff}}(\psi)}{d\psi}
    =-D(\psi)\frac{d(\beta\hat{E}^{\mathrm{eff}}(\psi))}{d\psi}
\end{equation}
where
\begin{equation}\label{PotEff2}
    \beta\hat{E}^{\mathrm{eff}}(\psi)=-\int^{\psi}
    \frac{v(\psi')}{D(\psi')}d\psi'+const.
\end{equation}
with $\beta\hat{E}^{\mathrm{eff}}$ differing from $\beta
E^{\mathrm{eff}}$ by the state-dependent contribution $ln
D(\psi)$.
For such systems with state-dependent noise we require (local)
estimates of {\em both} drift and diffusion coefficients for
effective potential stepping.
These can be used in eq.(\ref{PotEff}) (resp. eq.(\ref{PotEff2}))
to compute (differences in) the true effective potential
$E^{\mathrm{eff}}$ (resp. the ``auxiliary" effective potential
$\hat{E}^{\mathrm{eff}}$)
\begin{equation}\label{EqnMod2}
    \frac{d(\beta E^{\mathrm{eff}}(\psi))}{d\psi}=-\frac{1}{D(\psi)}\frac{d<\psi>}{dt}
    +\frac{1}{D(\psi)}\frac{dD(\psi)}{d\psi}.
\end{equation}

When temperature is not part of the problem description one
considers the SDE

\begin{equation}\label{SDEgen}
    d\psi=A(\psi)dt+B(\psi)dW_{t}
\end{equation}
which has the following stationary distribution

\begin{equation}\label{SDEgenStat}
    P_{st}(\psi)=\frac{c}{B(\psi)}e^{\int^{\psi}\frac{A(\psi')}{B(\psi')}d\psi'},
\end{equation}
$c$ being a normalization constant chosen such that
$\int_{-\infty}^{\infty} P_{st}(\psi')d\psi'=1$.
Local estimates of $A(\psi)$ and $B(\psi)$ can then, in a similar
approach as above, be used to step backwards in effective
potential.

\clearpage
\bibliographystyle{TF1}
\bibliography{RINGpaperBIB}             

\clearpage
\newpage
{\rm Figure 1. }{Schematic of reverse projective integration. The
thick gray line indicates the position on the slow manifold as a
function of time on a forward trajectory. The solid circles are
configurations along microscopic trajectories run forward in time,
as indicated by the short solid arrows. The long dashed arrows
indicate the reverse projective steps, which result in an
initialization near, but slightly off, the slow manifold.} \vskip
4 mm
 {\rm Figure 2. }{Schematic of forward and backward
stepping of ring nodes (light circles) in time on an energy
landscape in the vicinity of  fixed point (dark circles). Solid
lines are energy contours, dashed lines connect ring nodes at each
step, and arrows indicate the direction of the ring evolution.}
\vskip 4 mm
 {\rm Figure 3. }{Contour map of the
M\"{u}ller-Brown Potential for $-1<x<1,-0.5<y<1$. Contour lines
are shown in black (white) for $V(x,y)<0$ ($V(x,y)>0$). Stationary
points of the potential, their classification and energy are
tabulated for the region illustrated.} \vskip 4 mm

 {\rm Figure 4.
}{Distribution of nodes produced by integration of
equation~(\ref{eq:ringinit}) with initial condition above (white
nodes and contour lines) and below (black nodes and
  contour lines) the saddle point energy. Below the saddle point there is a separation
  of isopotential contours in each well -- the saddle point
  isopotential contours ``split'' in two.}
\vskip 4 mm {\rm Figure 5. }{Stages of ring evolution: backward
stepping (in time $\Delta t$,
  arc-length $\Delta s$, or potential $\Delta V$), followed by nodal redistribution.}
\vskip 4 mm {\rm Figure 6. }{Reverse time stepping on
M\"{u}ller-Brown Potential with $\Delta t=5\times10^{-5},
  N=80$ (successive rings are shown at intervals of 10 steps and arrows
   indicate direction of ring evolution).}
\vskip 4 mm {\rm Figure 7. }{Arc length stepping on
M\"{u}ller-Brown Potential with $\Delta s=0.01,  N=80$ (successive
rings are shown at intervals of 10 steps and arrows indicate
direction of ring evolution).} \vskip 4 mm {\rm Figure 8. }{Energy
stepping in a smooth, asymmetric 1D energy well.} \vskip 4 mm {\rm
Figure 9. }{Potential stepping on M\"{u}ller-Brown Potential with
$\Delta V=1.45, N=80$  (successive rings are shown at intervals of
10 steps and arrows indicate direction of ring evolution).} \vskip
4 mm {\rm Figure 10. }{Potential stepping on the M\"{u}ller-Brown
Potential with $\Delta V=0.75, N=160$
 (successive (colored) rings are shown at intervals of 10 steps).
Successive rings obtained by reverse  integration starting from
each of the minima on the landscape are shown.} \vskip 4 mm {\rm
Figure 11. }{Potential stepping on M\"{u}ller-Brown Potential with
$\Delta
 V=0.75, N=160$  (successive (colored) rings are shown at intervals of
 30 steps). Positions (red circles) of the minimum in
 gradient norm along the ring are shown at intervals of
 5 steps in ring integration. Three different viewpoints of the
same ring evolution are shown. Black arrows indicate the direction
of ring evolution out of each minimum. Top row: 3D view; bottom
row: 2D overhead view (gray arrows indicate position of views
shown in top row).} \vskip 4 mm {\rm Figure 12. }{Left: 100 rounds
of potential-stepping ring evolution   using coarse Euler and an
inner SDE integrator with $\Delta V=5\times10^{-2}, D=1, N=200$;
   a redistribution that concentrates
  nodes in regions of largest ring curvature is performed
  every 10 reverse ring integration steps.
The ring is initially centered at
  $(-1,-1)$. 50 replica runs are performed with the SDE integrator, each run
 for $t_{tot}=0.5$ (with time step size $\Delta t=2.5\times 10^{-3}$),
for drift estimation at each node. Contours of the function
$V(x,y)$ (defined in equation~(\ref{eq:SDE_V(xy)})) are shown.
Ring nodes are shown for every step. Right: the effective
potential associated with each ring node is shown (indicated by
color) computed using local drifts and diffusions using
(\ref{eq:lineint1}) -- it is plotted superimposed on the landscape
of the potential (\ref{eq:SDE_V(xy)}). Evolving ring nodes with
$x<-1.2$ are omitted for clarity. Points on a single
representative effective potential contour (using reverse
drift-based integration) are plotted as black symbols in the
$V(x,y)=-10$ plane at the base of the figure; points along the
actual potential contour are shown as red symbols.} \vskip 4 mm
{\rm Figure 13. }{Left: 100 rounds of drift potential-stepping
ring  evolution using an explicit tau-leaping inner Gillespie
simulator with   $N=200$; nodal redistribution is performed every
10 reverse ring integration (coarse Euler) steps.  The ring is
initially centered at
  $(-1,-1)$. 50 replica Gillespie simulation runs are performed, each
  run with $10,000$ particles and the explicit tau leaping parameter $\epsilon=0.03$.
    For the reverse integration $\Delta V=5\times10^{-2}$.
Contours of the function $V(x,y)$ (defined in
equation~(\ref{eq:SDE_V(xy)})) are shown. Right: 3D-view of
reverse ring  integration shown left with estimated potential of
each node shown in color. Colorbar indicates value of $V(x,y)$.
Evolving ring nodes with $x<-1.2$ are omitted for clarity.
 Contours of the function $V(x,y)$ (defined in equation~(\ref{eq:SDE_V(xy)}))
  are shown in 3D. Points
on a single representative potential contour (as computed using
reverse integration) are plotted as black symbols in the
$V(x,y)=-10$ plane at the base of the figure; points along the
actual potential contour are shown as red symbols.} \vskip 4 mm
{\rm Figure 14. }{Free energy landscape for the alanine dipeptide
in the
  $\phi-\psi$ plane (1$k_BT$ contour lines). Structures are shown
    corresponding to the right-handed $\alpha$-helical minimum (left), the
  extended minimum (right), and the transition state between them (middle).}
\vskip 4 mm {\rm Figure 15. }{Alanine dipeptide ring integration.
Left panel: Extended structure minimum: 30 rounds of reverse
   (coarse Euler) ring   integration (number of ring nodes $N$=12) with scaled effective
  potential steps. Note that the scaled steps correspond to constant
   steps in free energy only if the effective
  diffusion tensor is diagonal with identical, constant entries,
   which appears to be a good approximation here.
Eigenvectors corresponding to  positive eigenvalues for candidate
saddle points determined from ring integration are shown (long red
arrows). Right panel: Downhill runs initialized at transition
regions suggested by the reverse ring integration from the
extended structure minimum. Initial conditions (black dots) are
generated by umbrella sampling at a target coarse point selected
by perturbation along the unstable eigenvector at the saddle.
Final conditions for these downhill runs (purple dots) suggest
starting points for a new round of reverse integration from the
adjacent minimum. ($1k_BT$ contour lines used in both plots). Note
that both wells are plotted rotated by 90 degrees relative to
Figure~14.}
\vskip 4 mm {\rm Figure 16. }{Alanine dipeptide in water: 30
rounds of coarse reverse ring evolution (number of ring nodes
$N$=12, $D\beta\Delta E^{\mathrm{eff}}=0.05k_BT$)
 initialized in the neighborhood of both the right-handed $\alpha$-helical minimum
  (bottom ring), and the extended minimum (top ring). Rings (grey
  lines) connecting nodes (black solid circles) are shown. Colored
  energy contours are plotted at increments of $1k_BT$.}

\clearpage
\newpage
\begin{figure}[htbp]
\centering
\includegraphics*[bb=0 0 1000 575,scale=0.3]{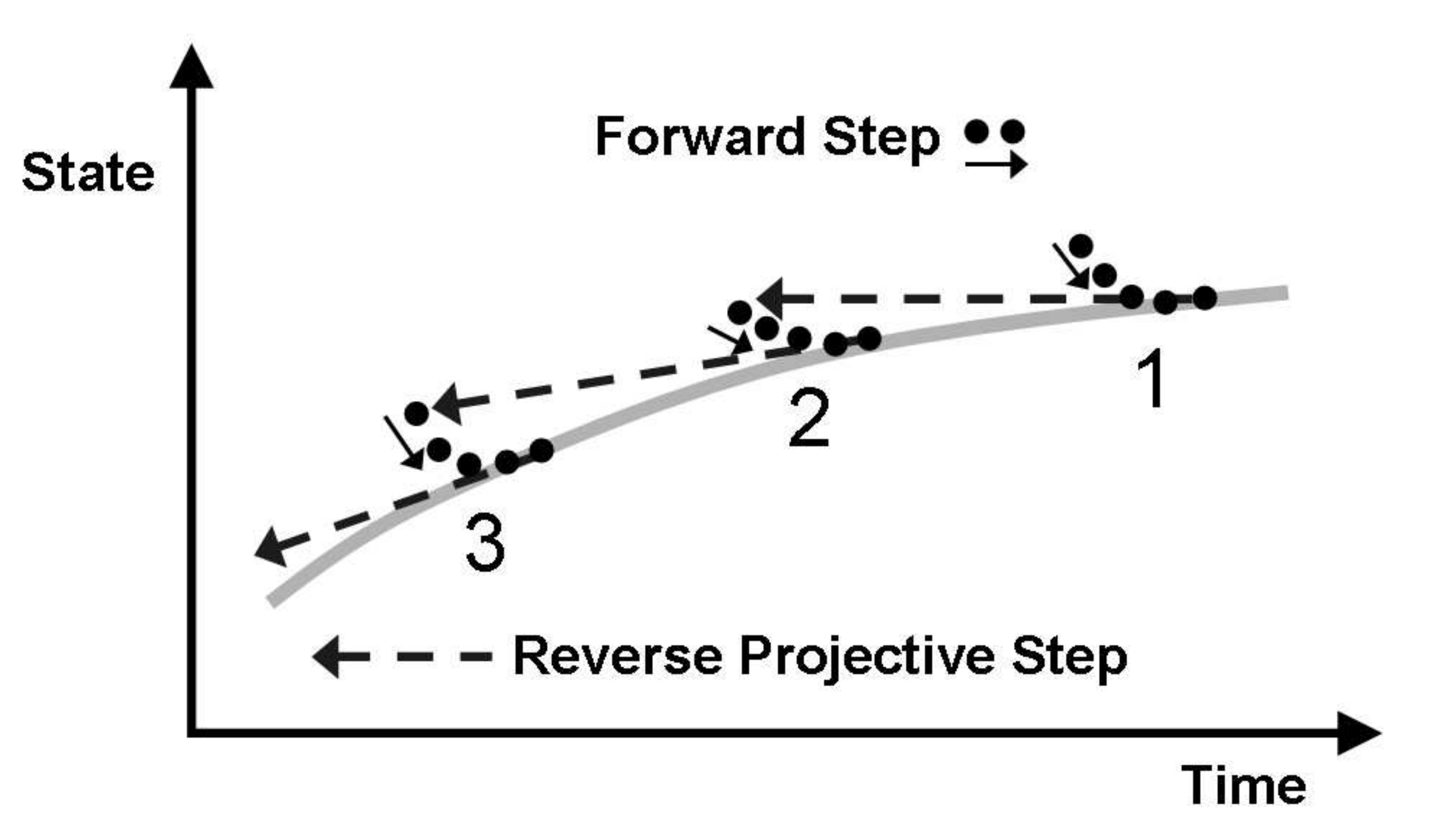}
\caption*{Figure 1}
\label{fig:reverseproj}
\end{figure}

\clearpage
\newpage
\begin{figure}[htbp]
\centering
\includegraphics*[bb=5 100 1475 850,scale=0.25]{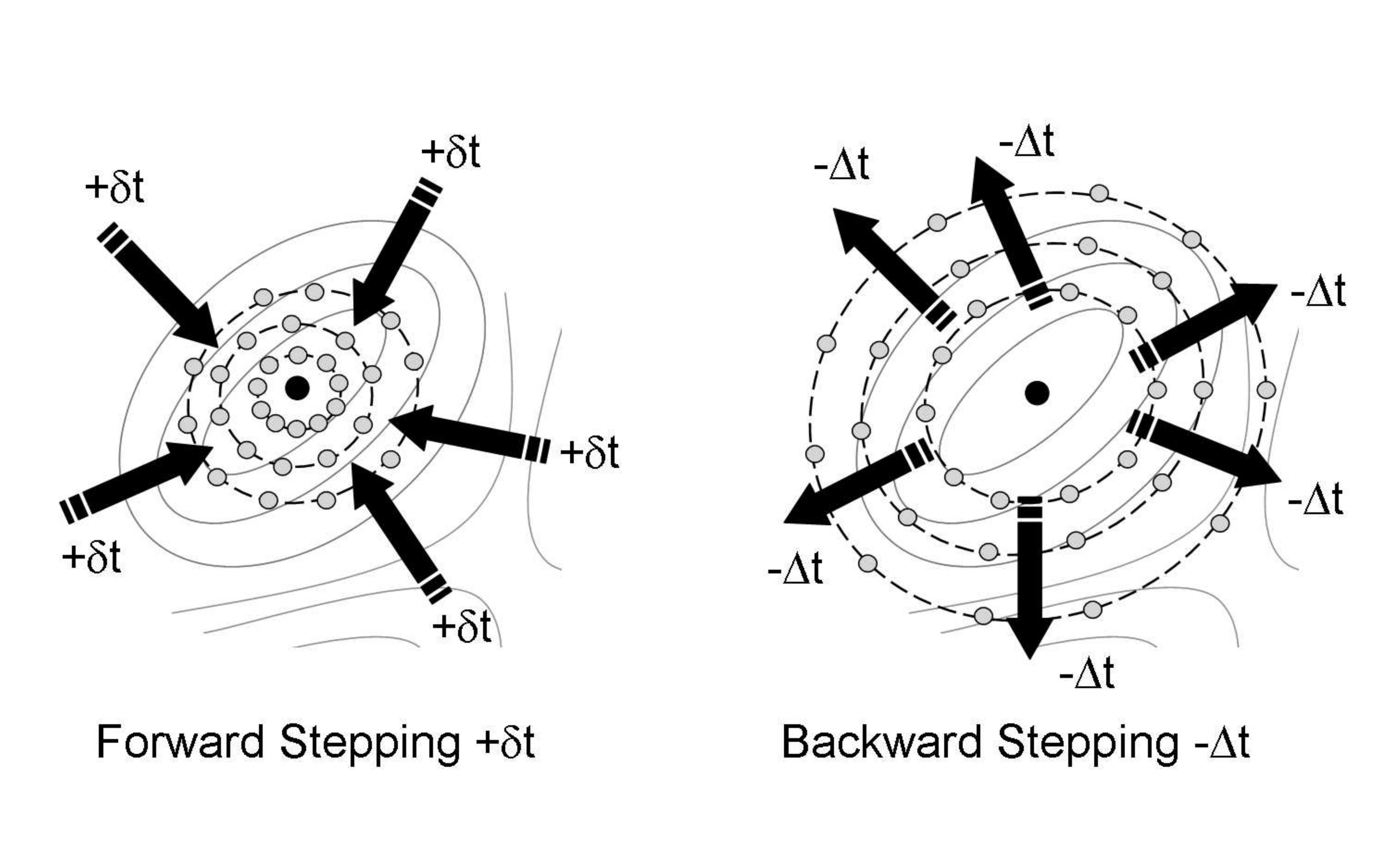}
\caption*{Figure 2}
\label{fig:for+back}
\end{figure}

\clearpage
\newpage
\begin{figure}[htbp]
\centering
\begin{minipage}{2.in}
\centering
\includegraphics[bb=120 160 550 650,scale=0.5]{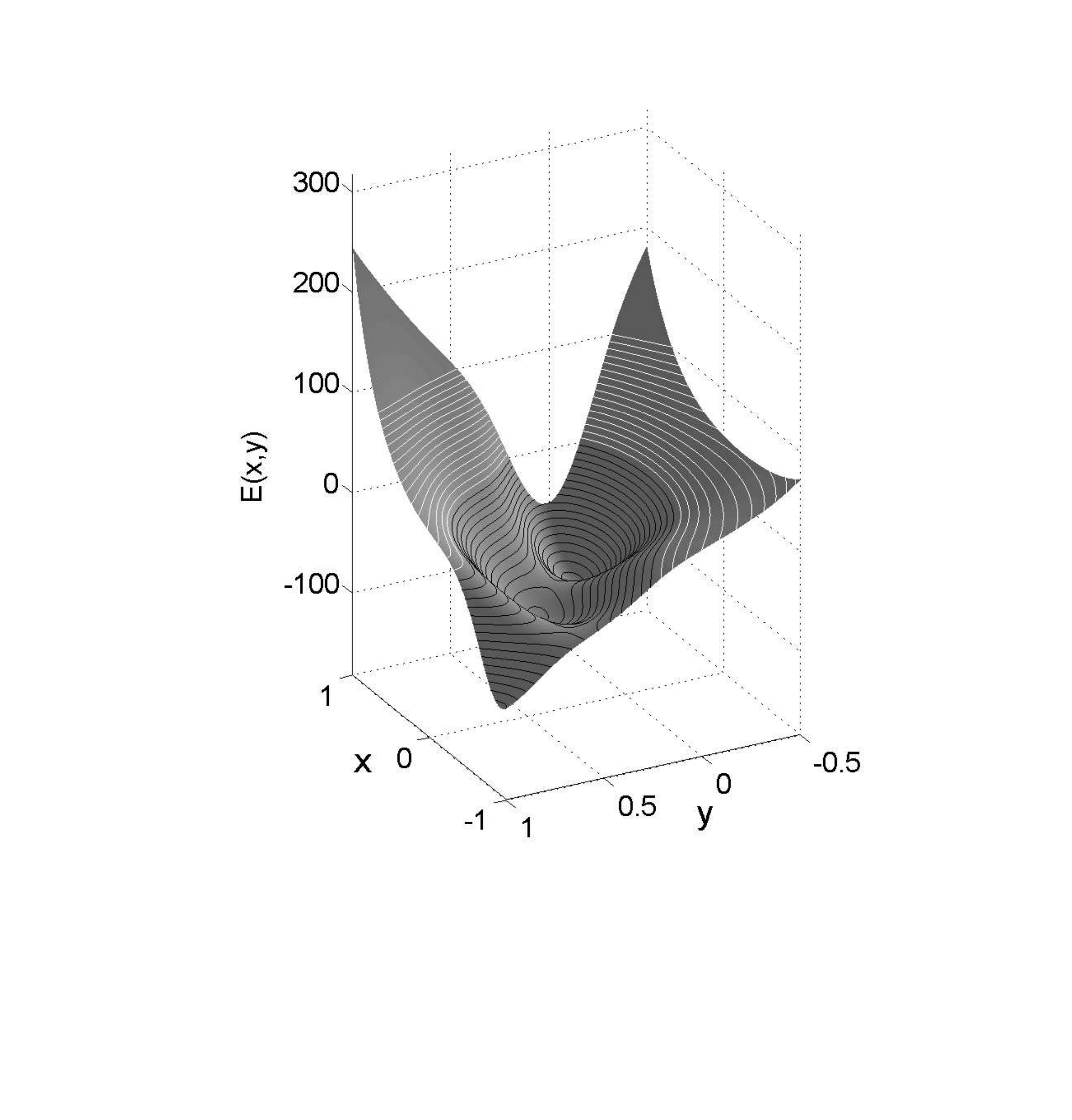}
\end{minipage}
\hspace{1.in}
\begin{minipage}{2.in}
\centering
\renewcommand{\arraystretch}{1.2}
\begin{tabular}{|c|c|c|c|}\hline
\emph{x}&\emph{y}&\emph{V(x,y)}&\emph{Feature}\\ \hline\hline
0.62    &0.03    &-108.2       &Minimum\\\hline
-0.05   &0.47&-80.8&Minimum\\\hline
-0.82   &0.62&-40.7&Saddle Point\\\hline
0.21   &0.29&-72.2&Saddle Point\\ \hline
\end{tabular}
\end{minipage}
\caption*{Figure 3}
\label{fig:MuellerPot}
\end{figure}

\clearpage
\newpage
\begin{figure}[htbp]
\hspace{0.2in}
\includegraphics*[bb=-20 185 620 775,scale=0.25]{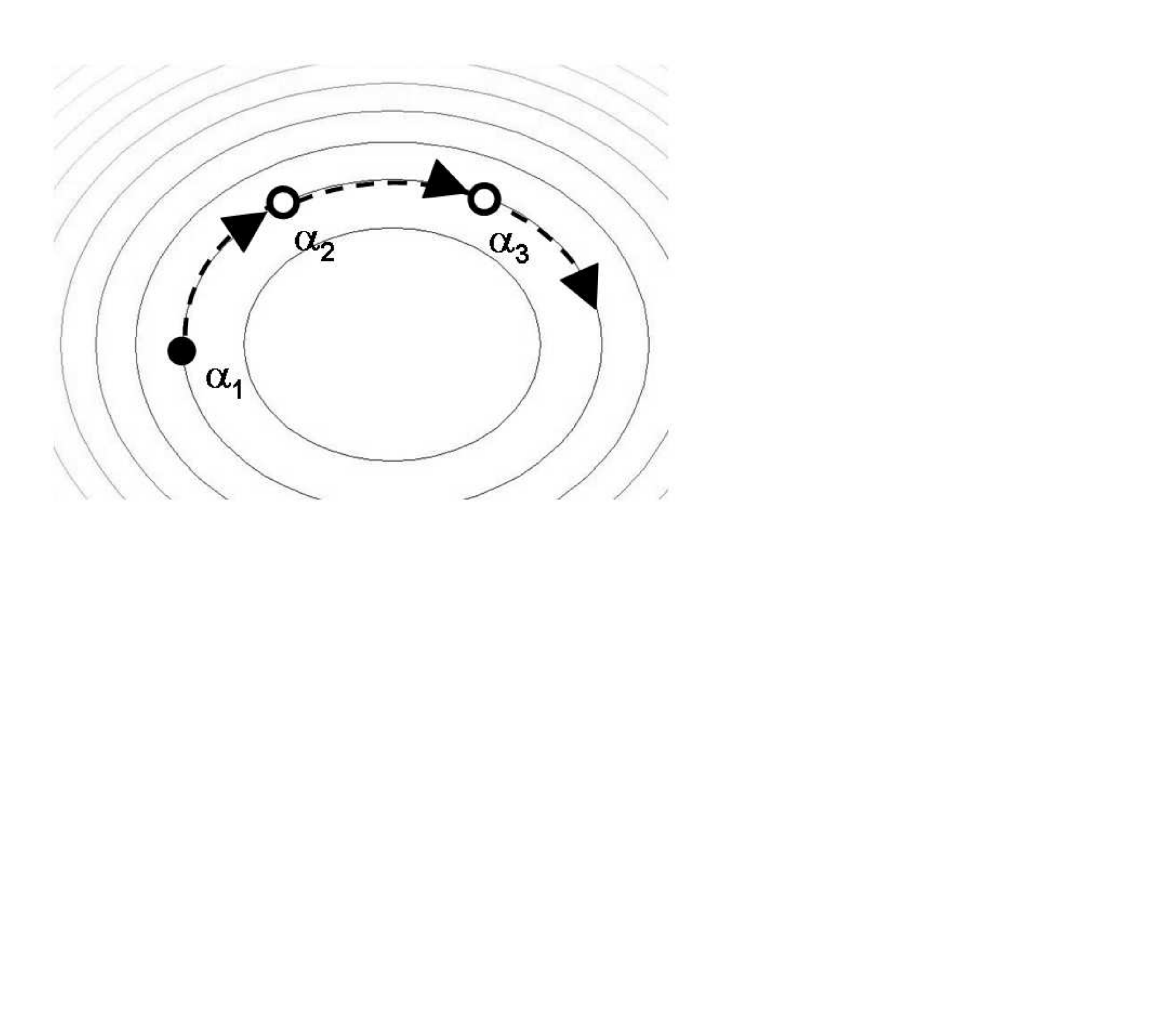}
\hspace{0.in}
\includegraphics*[bb=-20 -20 480 595,scale=0.55]{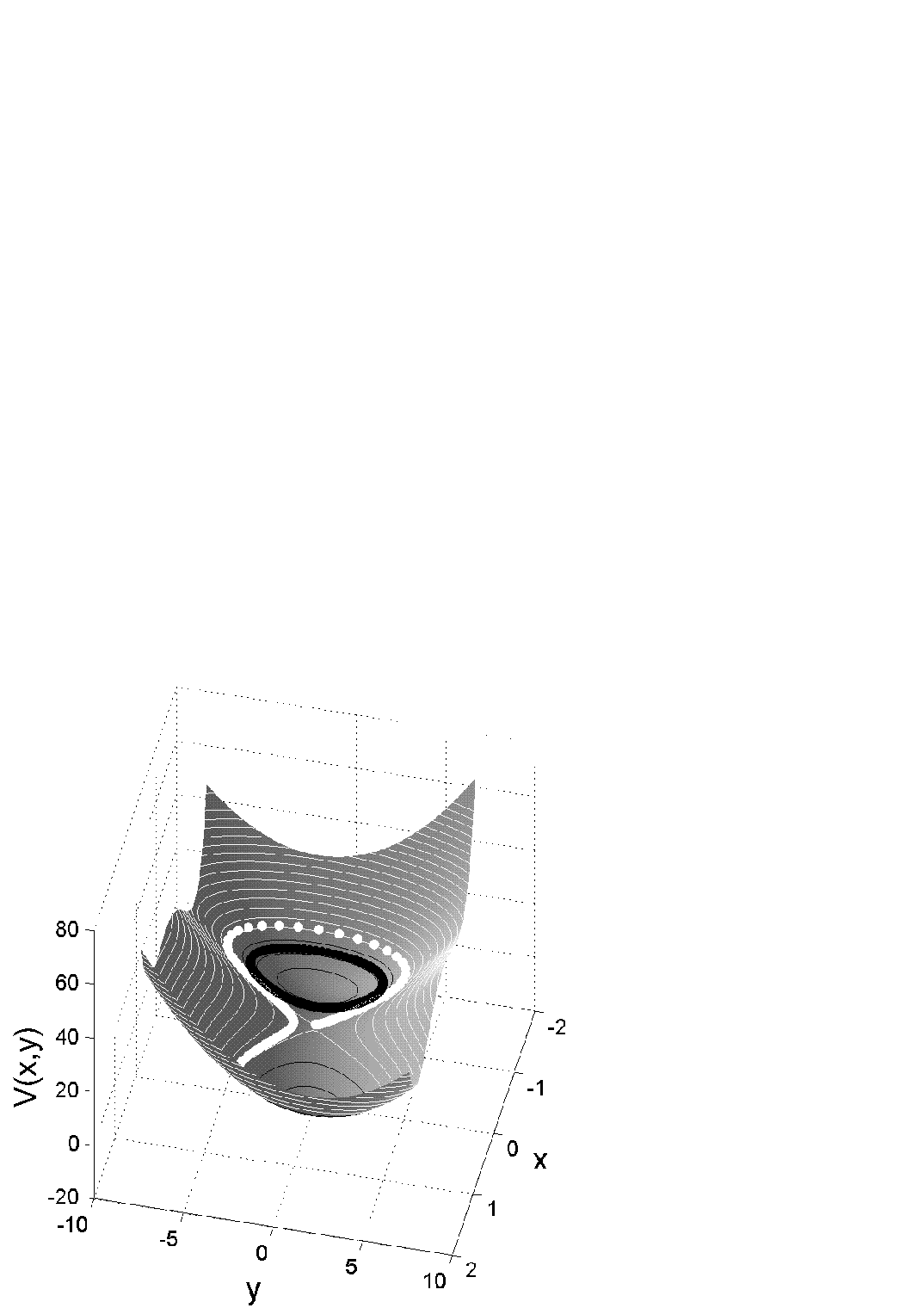}
\caption*{Figure 4}
\label{fig:ringinit}
\end{figure}

\clearpage
\newpage
\begin{figure}[htbp]
\centering
\includegraphics[bb=0 0 1400 465,scale=0.25]{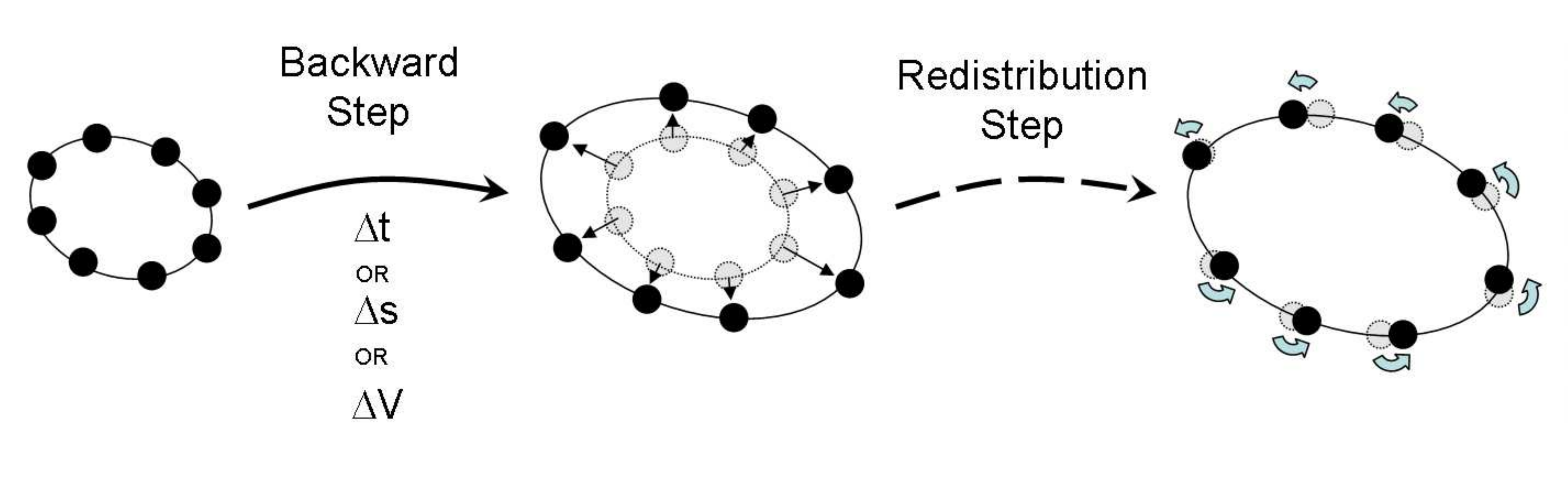}
\caption*{Figure 5}
\label{fig:ringevexp}
\end{figure}

\clearpage
\newpage
\begin{figure}[htbp!]
\centering
\includegraphics[bb=100 275 575 550,scale=0.475]{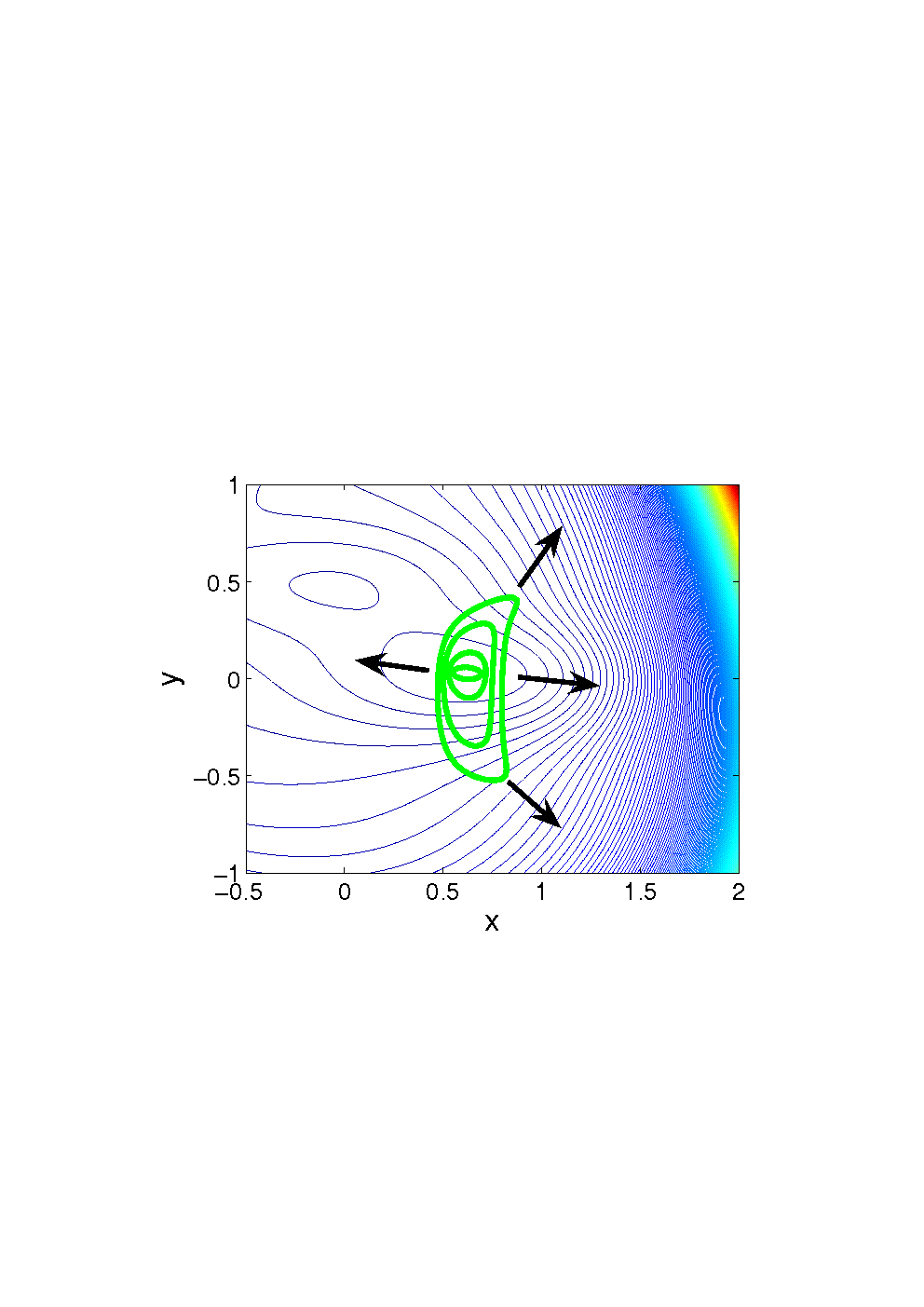}
\hspace{-0.2cm}
\includegraphics[bb=100 230 510 550,scale=0.55]{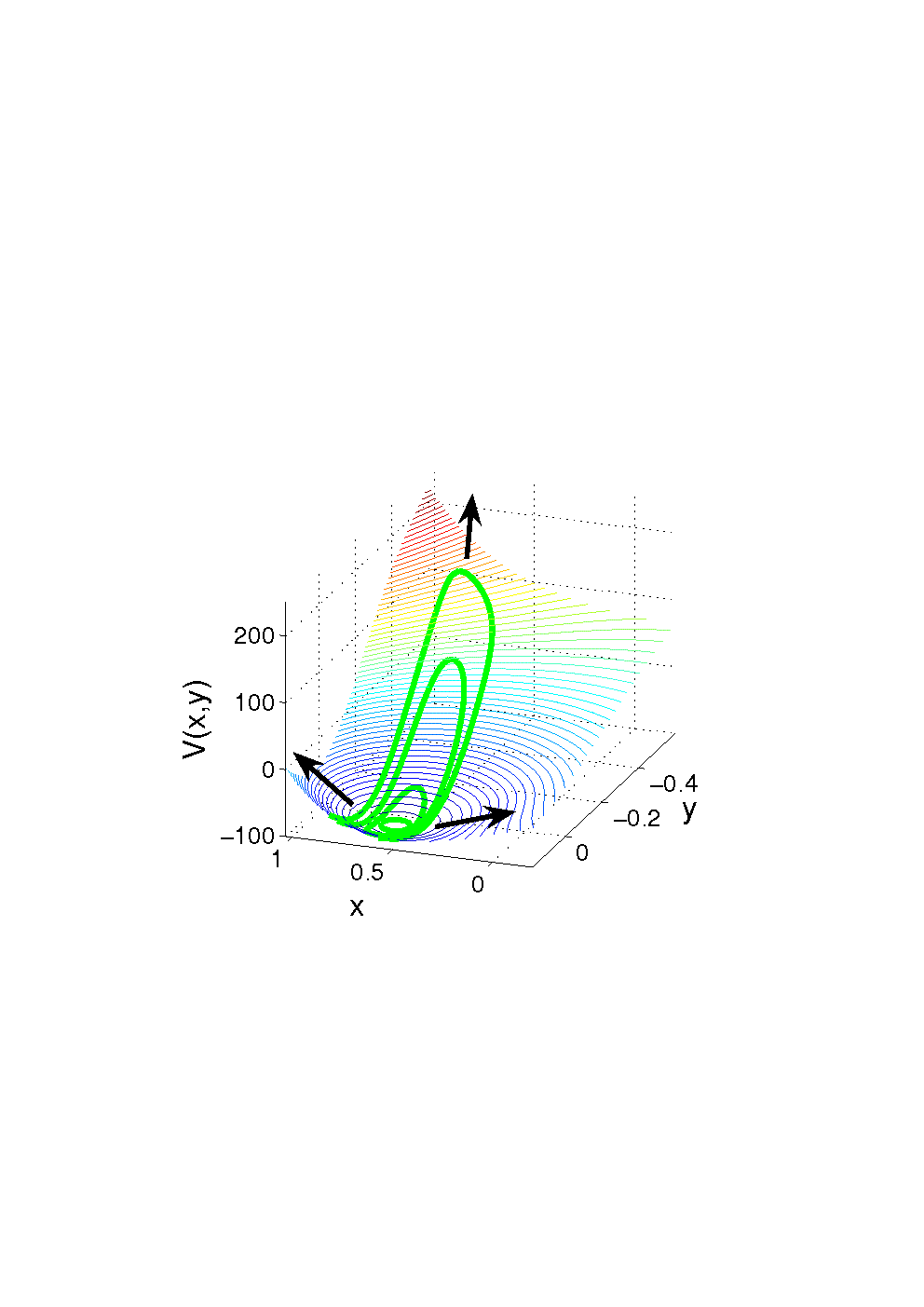}
\caption*{Figure 6}
\label{fig:Muellertime}
\end{figure}

\clearpage
\newpage
\begin{figure}[htbp!]
\centering
\includegraphics[bb=97 38 513 550,scale=0.45]{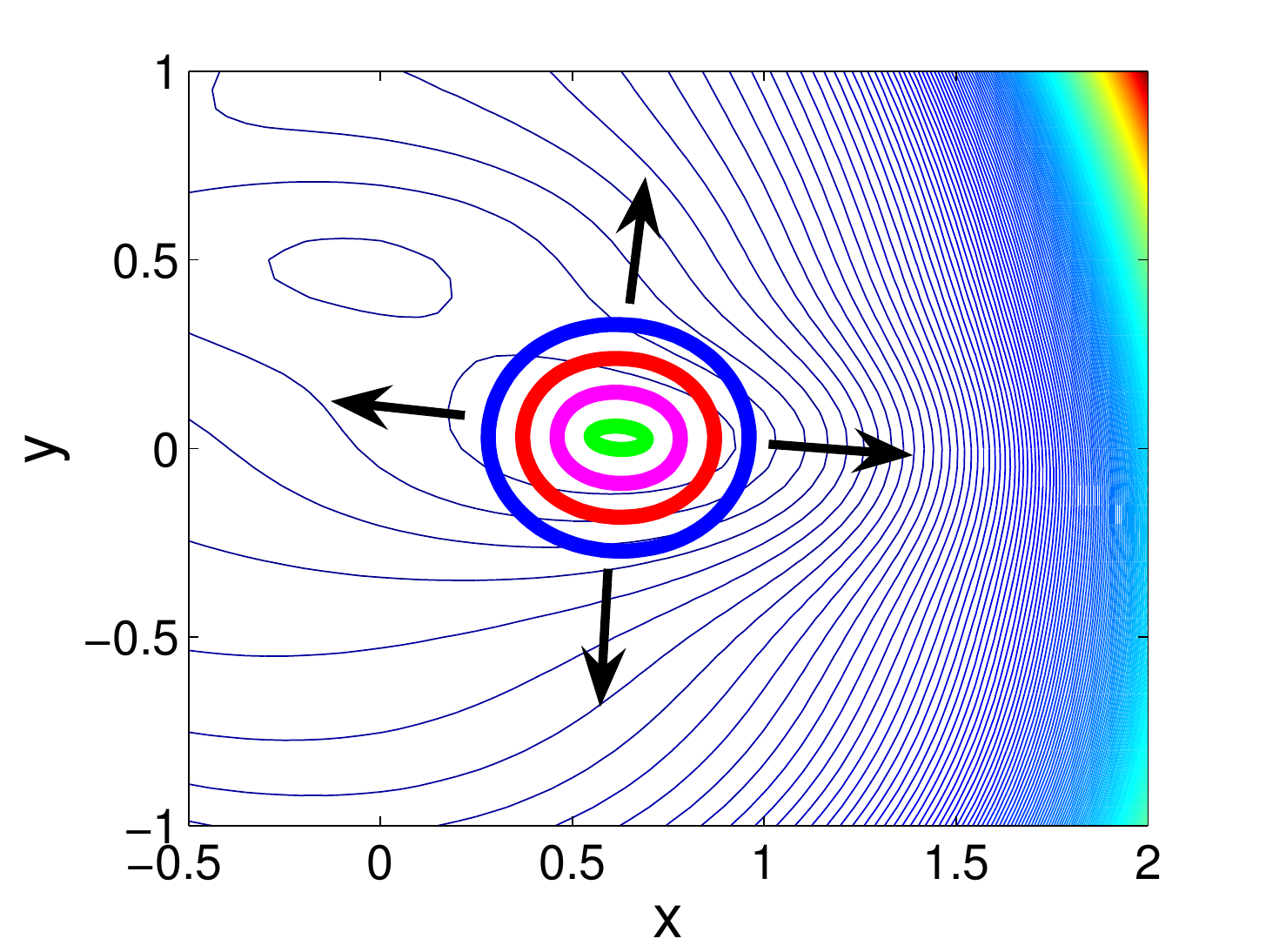}
\hspace{-0.2cm}
\includegraphics[bb=97 38 513 550,scale=0.55]{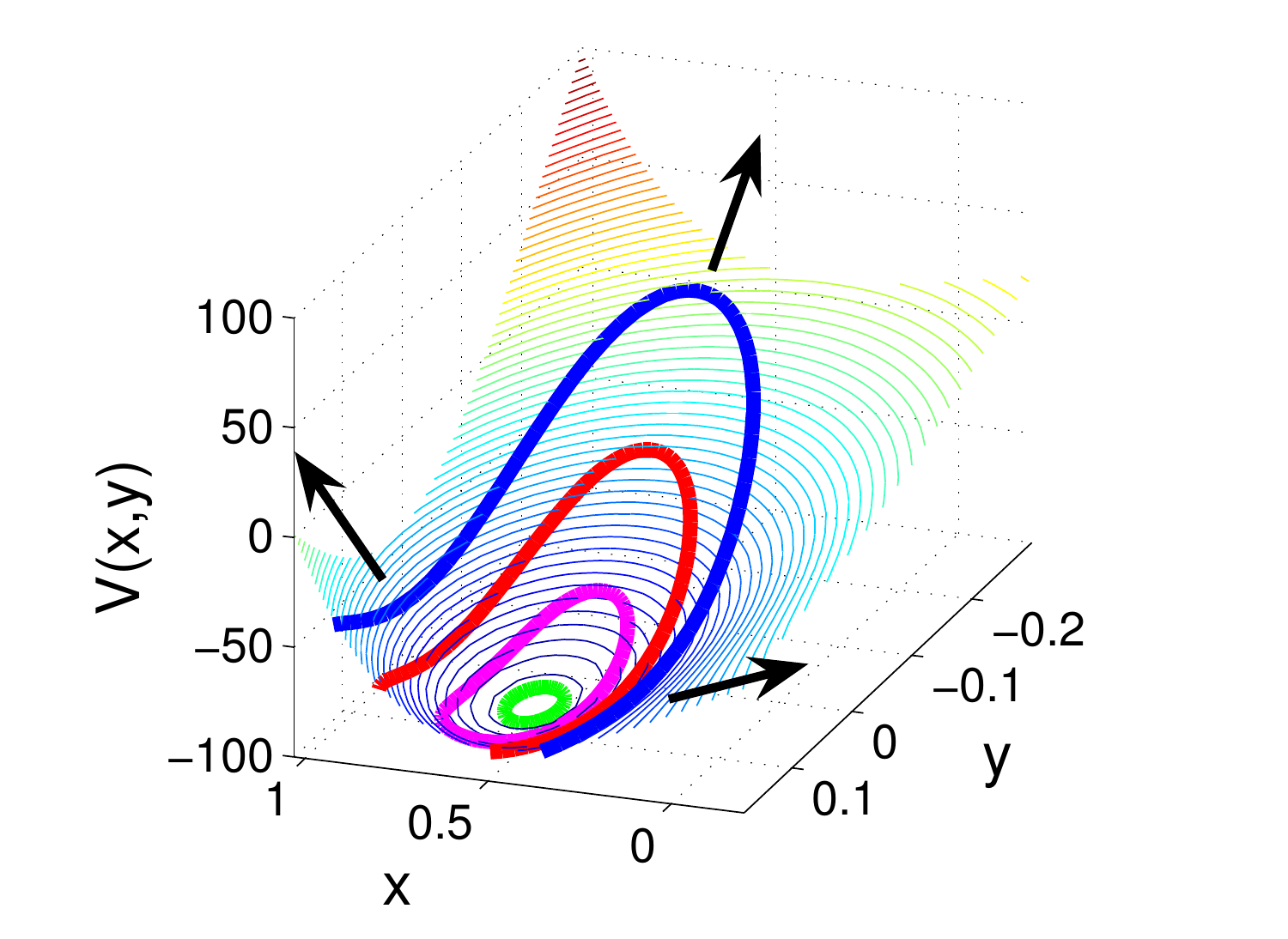}
\caption*{Figure 7}
\label{fig:Muellerarc}
\end{figure}

\clearpage
\newpage
\begin{figure}[htbp]
\centering
\includegraphics[bb=1 60 1146 461,scale=0.3]{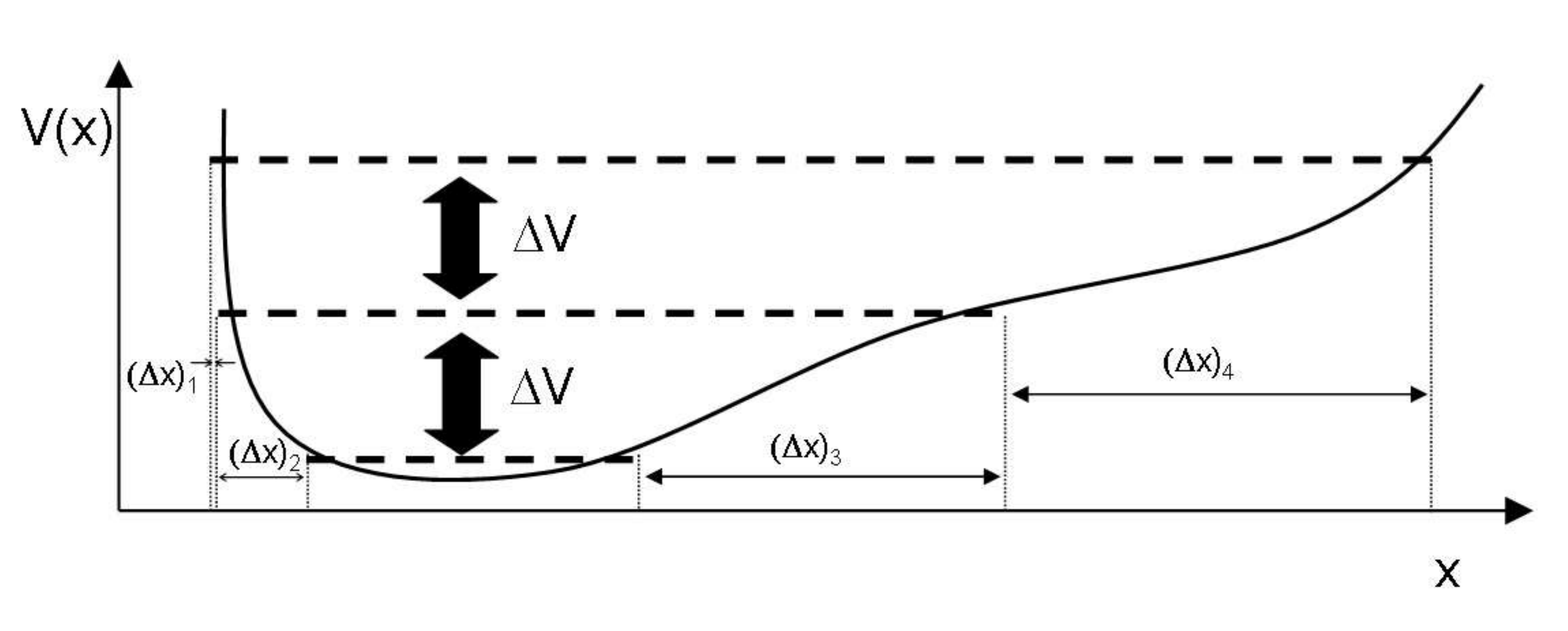}
\caption*{Figure 8}
\label{fig:estepscheme}
\end{figure}

\clearpage
\newpage
\begin{figure}[htbp!]
\centering
\includegraphics[bb=100 40 510 550,scale=0.5]{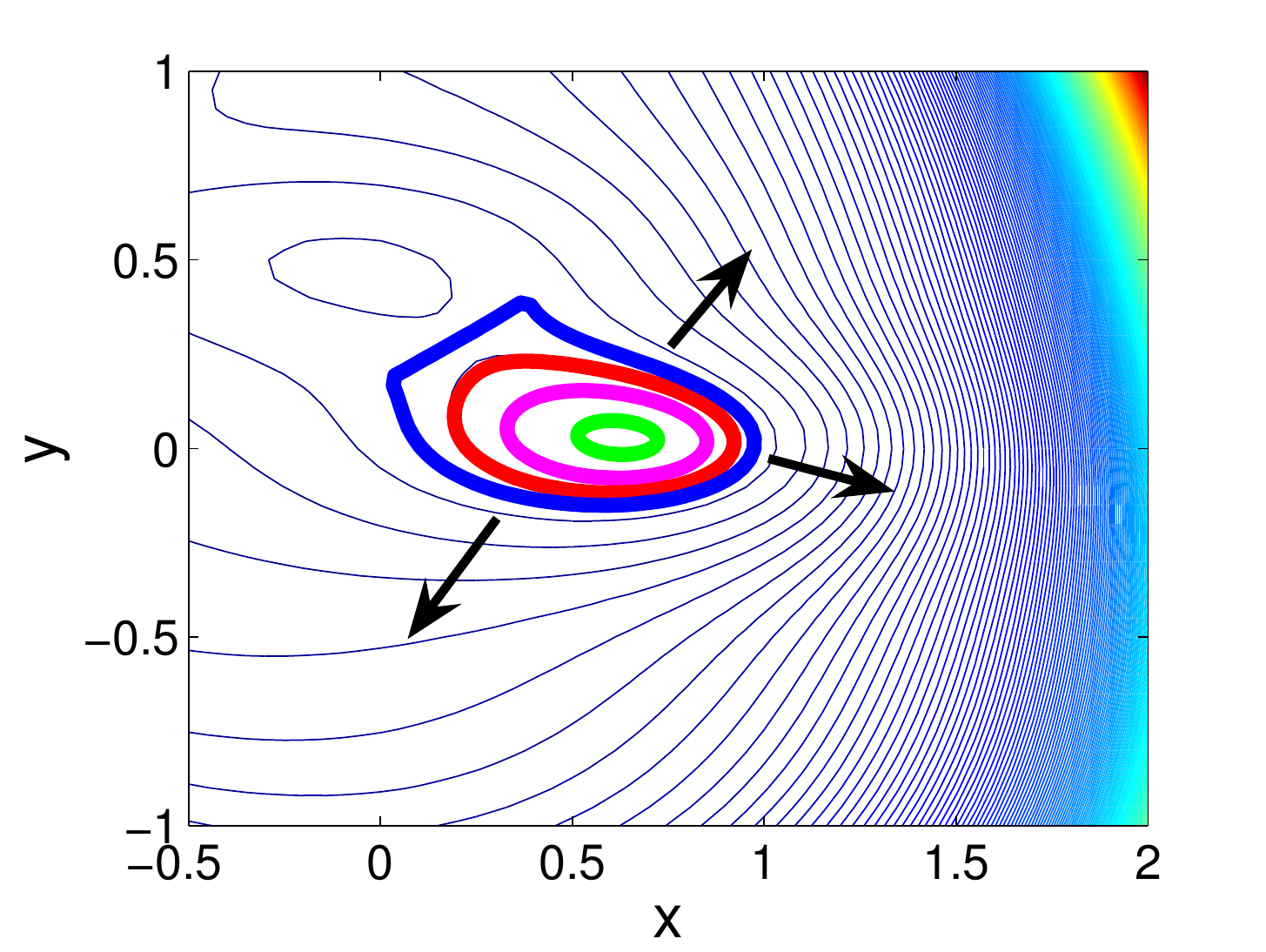}
\hspace{-0.2cm}
\includegraphics[bb=100 40 510 550,scale=0.55]{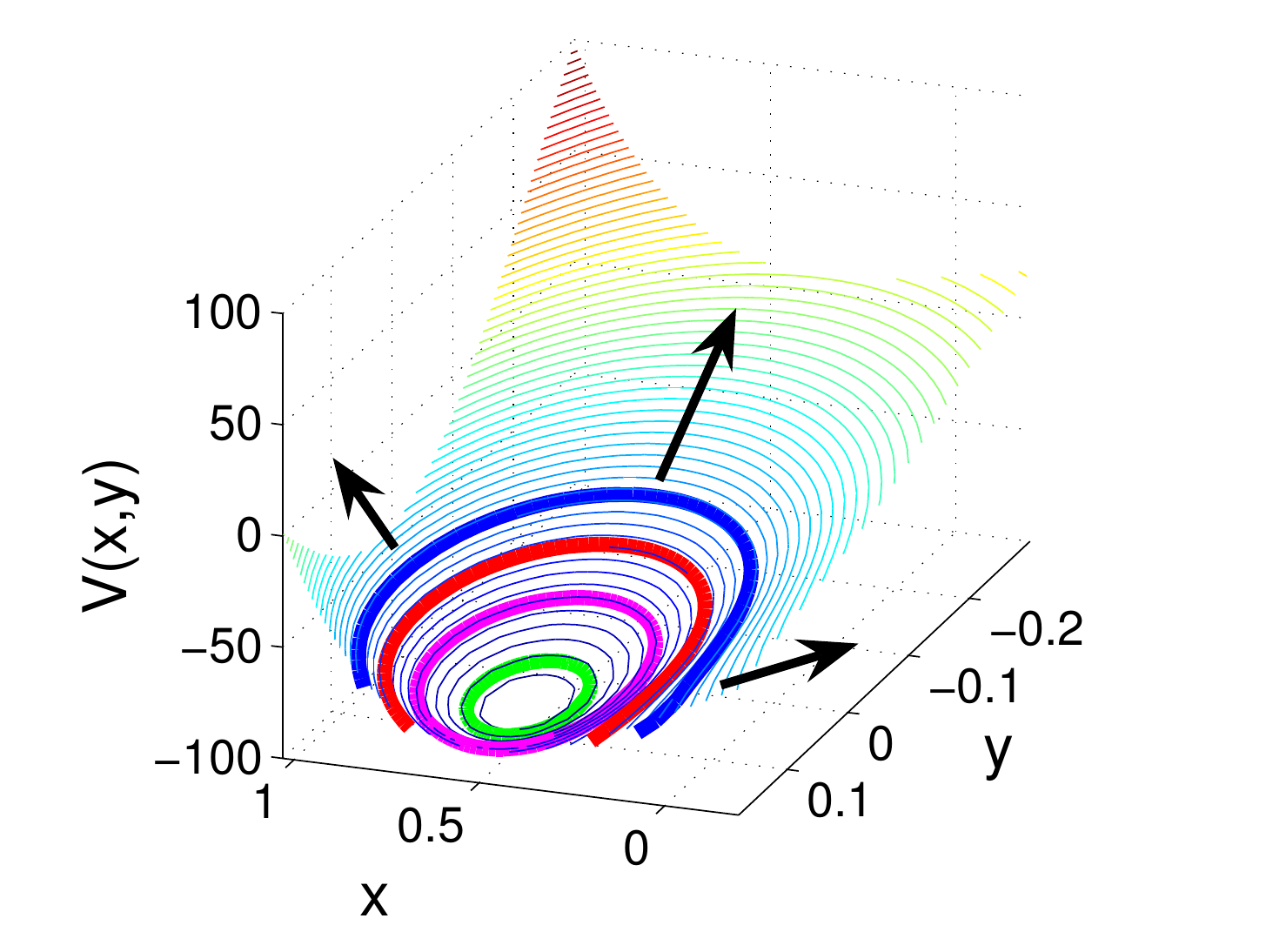}
\caption*{Figure 9}
\label{fig:Muellerenergy}
\end{figure}

\clearpage
\newpage
\begin{figure}[htbp]
\centering
\includegraphics[bb=20 350 540 490,scale=0.85]{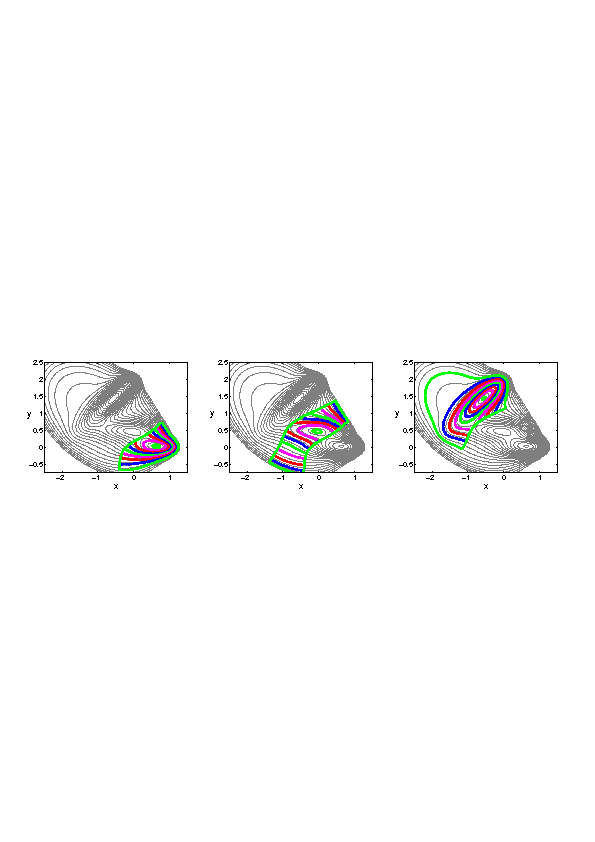}
\caption*{Figure 10}
\label{fig:MuellerenergyALL}
\end{figure}

\clearpage
\newpage
\begin{figure}[htbp]
\centering
\includegraphics[bb=155 300 435 500,scale=0.75]{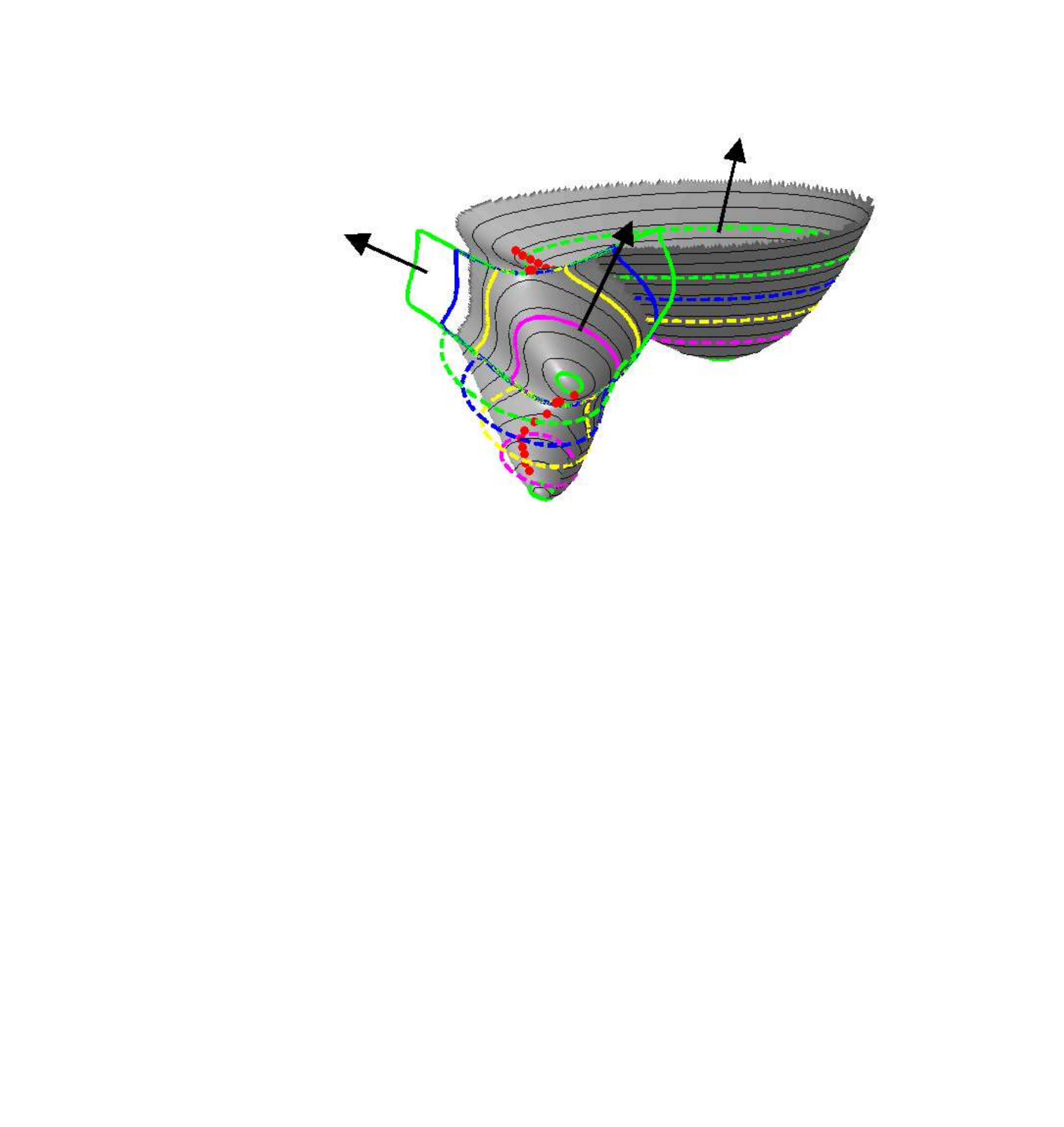}
\includegraphics[bb=110 330 415 540,scale=0.75]{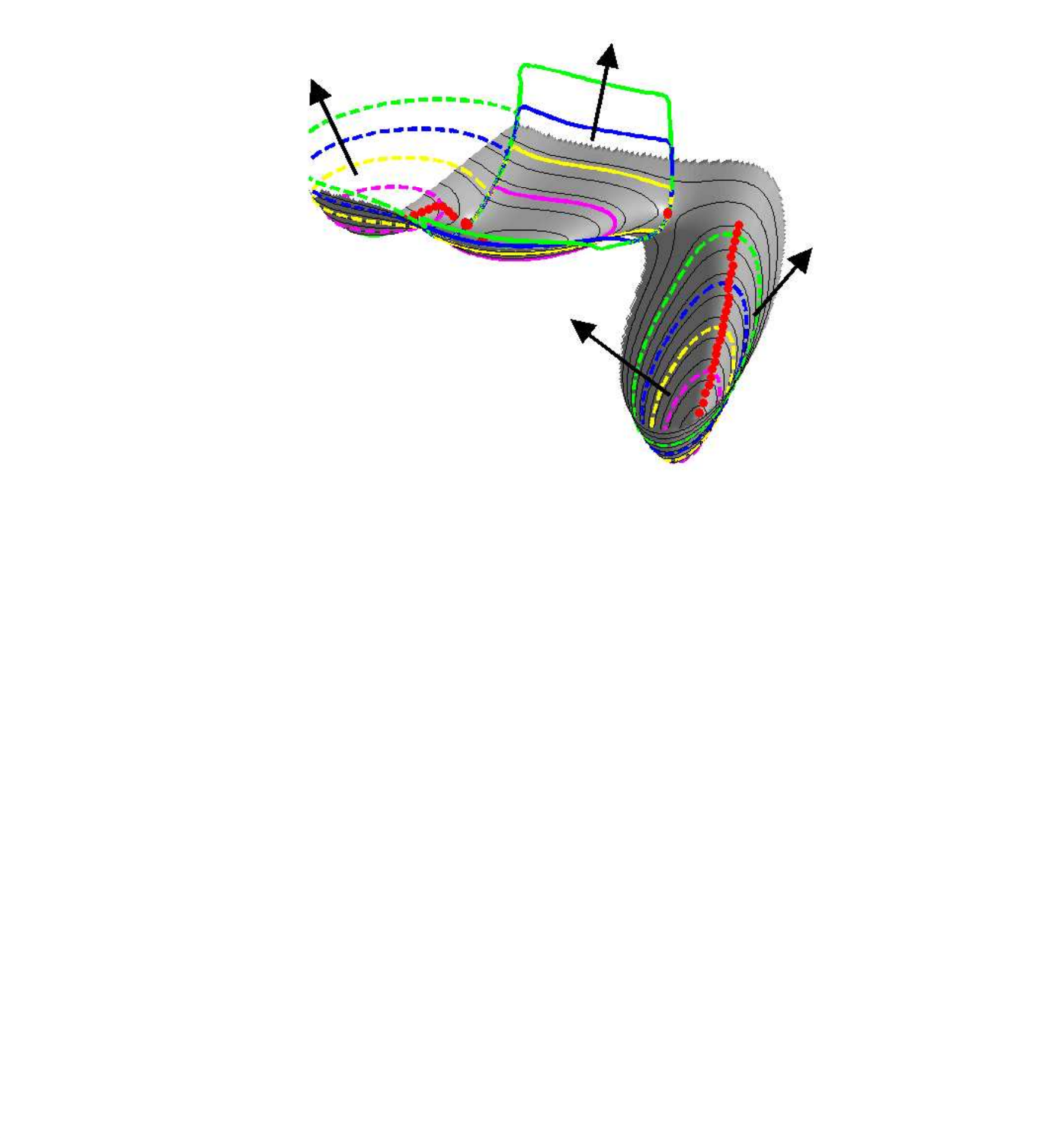}
\includegraphics[bb=65 185 510 575,scale=0.425]{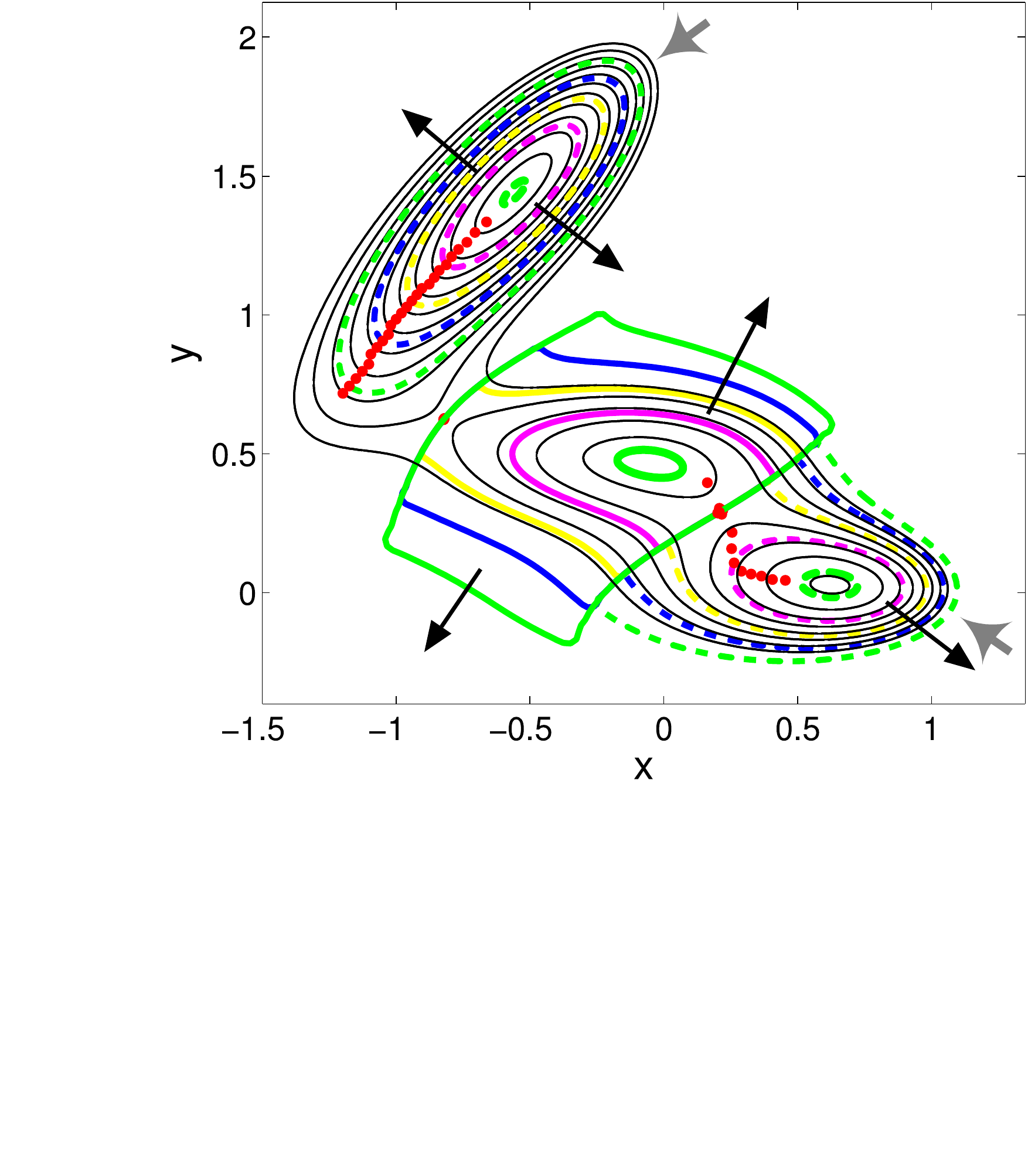}
\caption*{Figure 11}
\label{fig:MuellerenergyBASINS}
\end{figure}

\clearpage
\newpage
\begin{figure}[htbp]
  \centering
  \includegraphics[bb=25 150 575 625,scale=0.4]{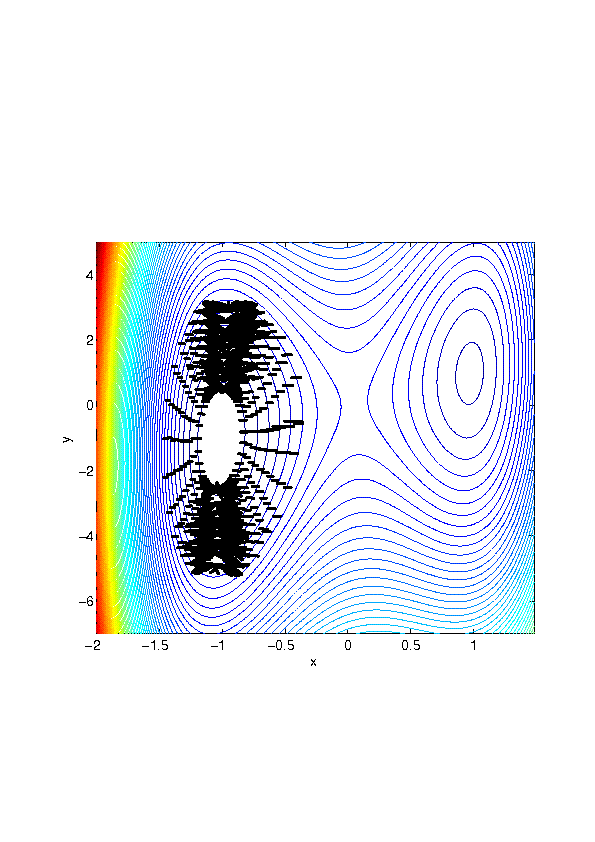}
  \hspace{0.2cm}
  \includegraphics[bb=25 50 570 620,scale=0.4]{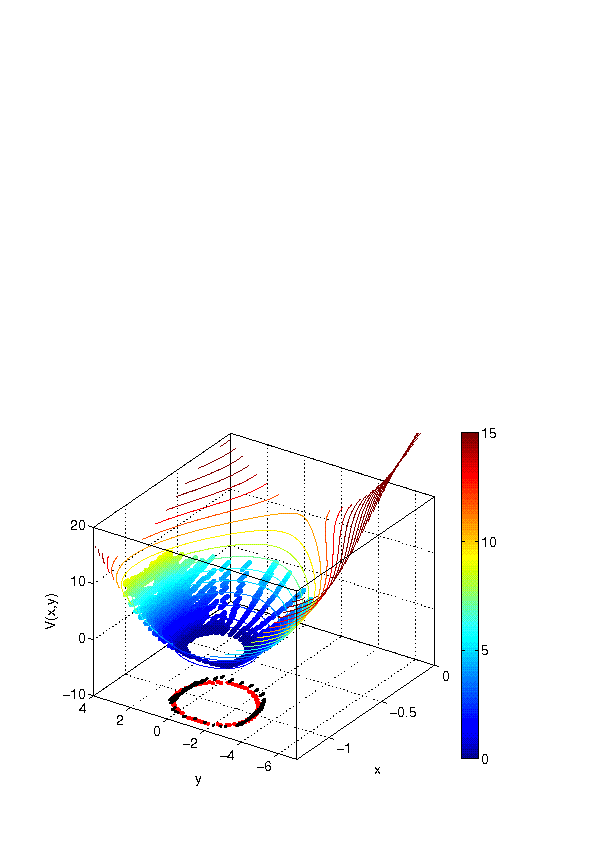}
\caption*{Figure 12}
\label{fig:SDE_ring}
\end{figure}

\clearpage
\newpage
\begin{figure}[htbp]
  \centering
  \includegraphics[bb=35 165 575 625,scale=0.4]{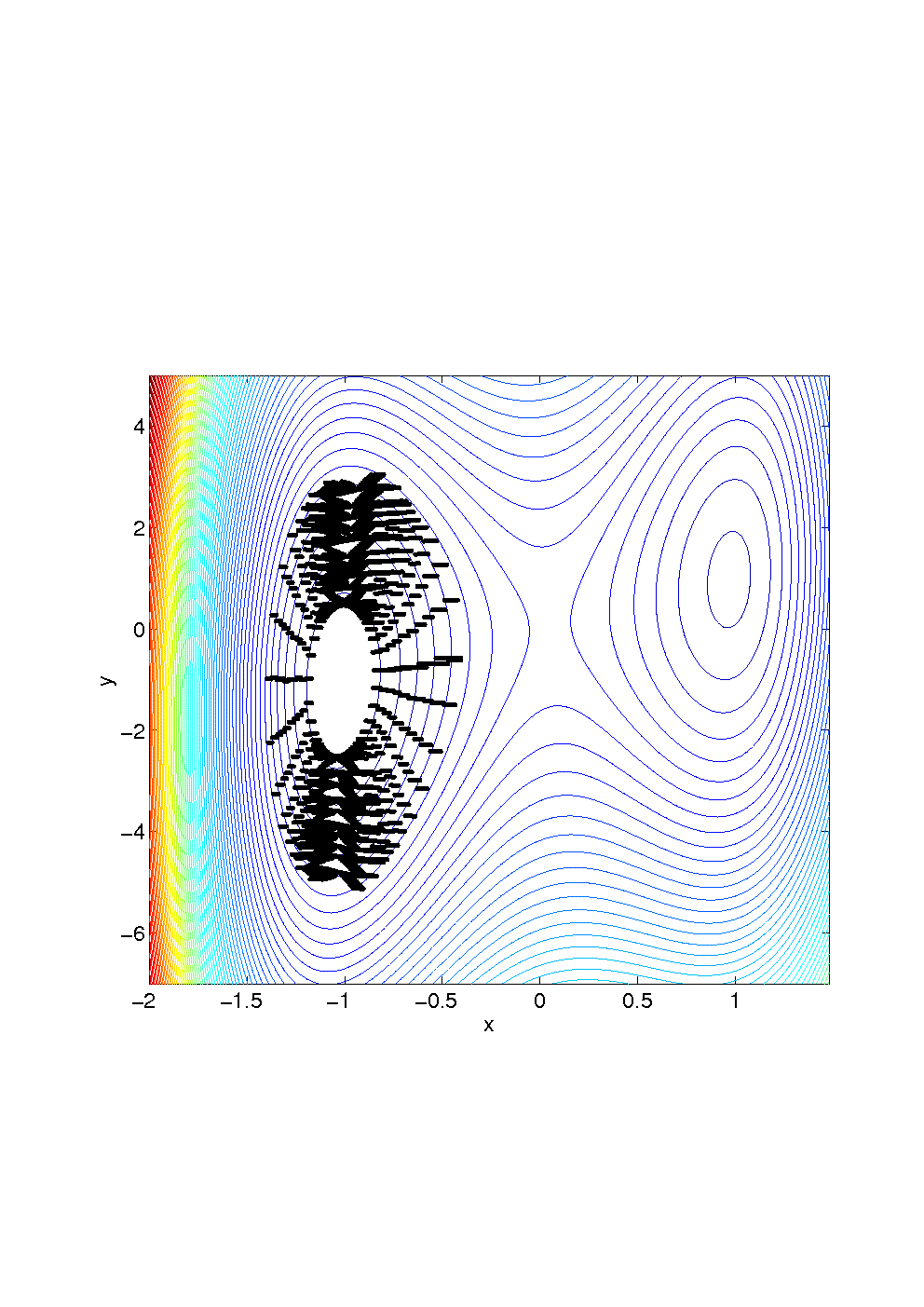}
  \hspace{0.2cm}
  \includegraphics[bb=55 10 580 630,scale=0.425]{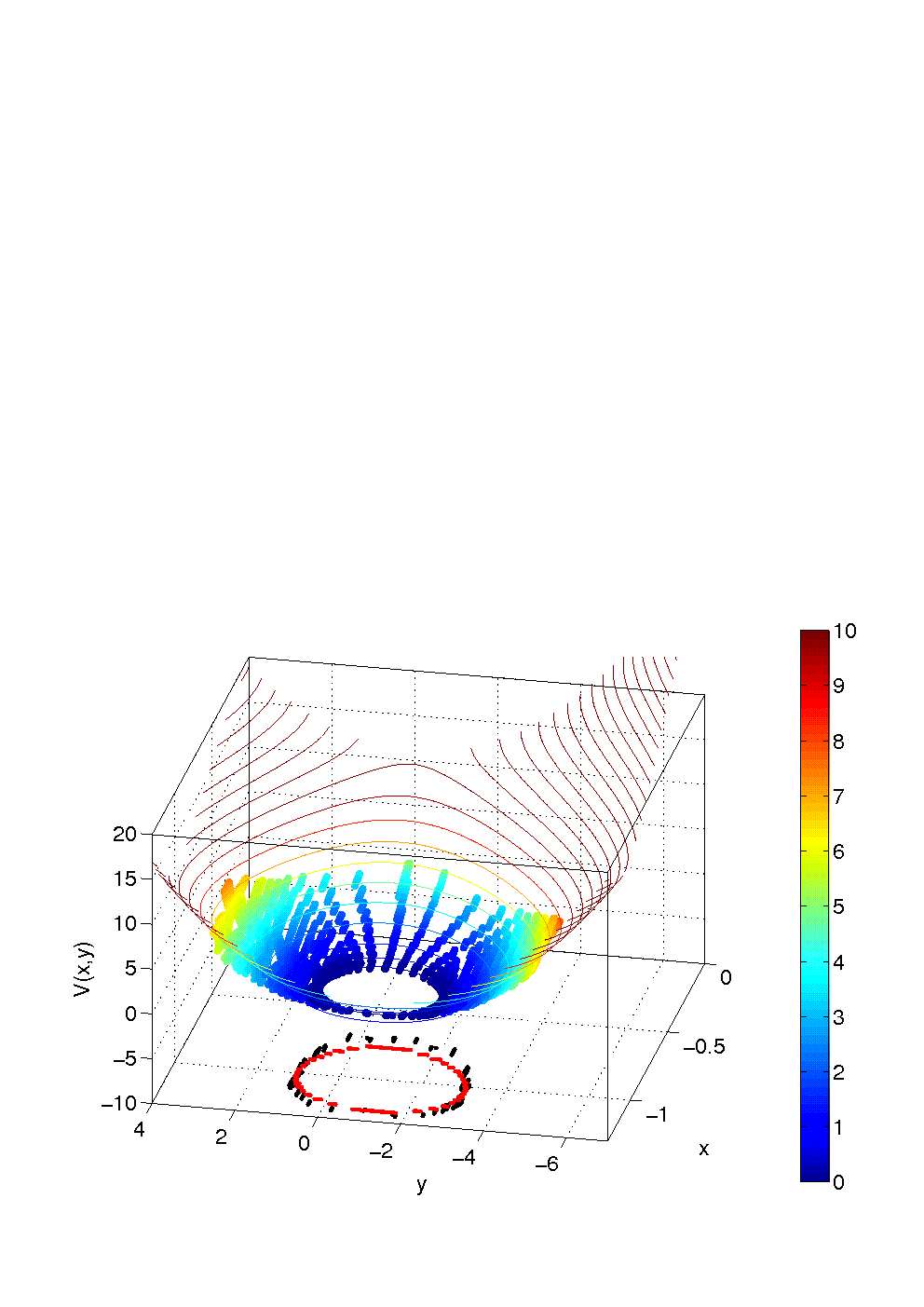}
\caption*{Figure 13}
\label{fig:Gillespie_ring}
\end{figure}

\clearpage
\newpage
\begin{figure}[htbp]
  \centering
  \includegraphics[bb=0 0 1250 1000,scale=0.475]{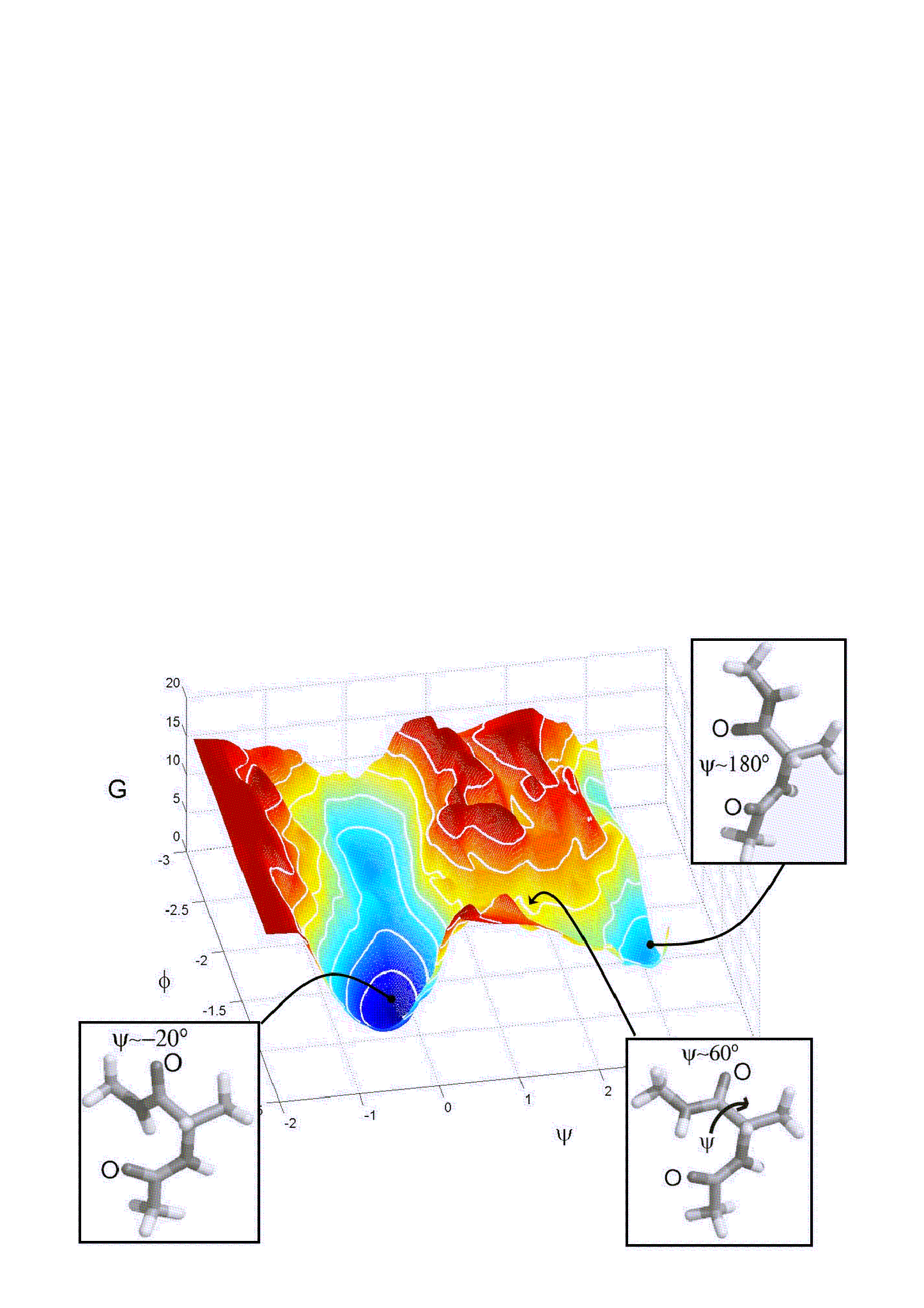}
\caption*{Figure 14}
\label{fig:aladip_landscape}
\end{figure}

\clearpage
\newpage
\begin{figure}[htbp]
 \centering
  \includegraphics[bb=0 0 810 660,scale=0.3]{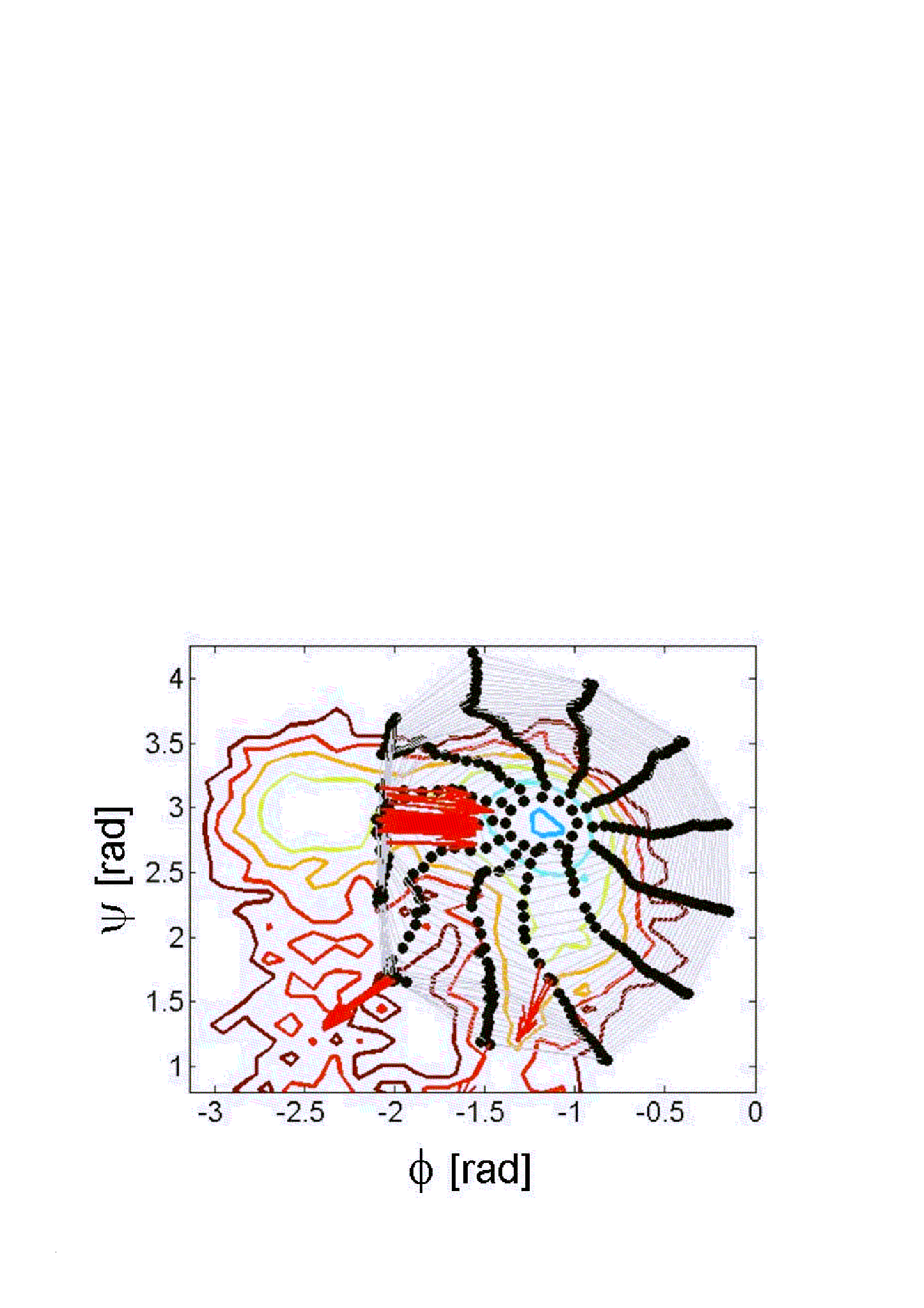}
  \hspace{-1.2cm}
  \includegraphics[bb=0 0 790 650,scale=0.3]{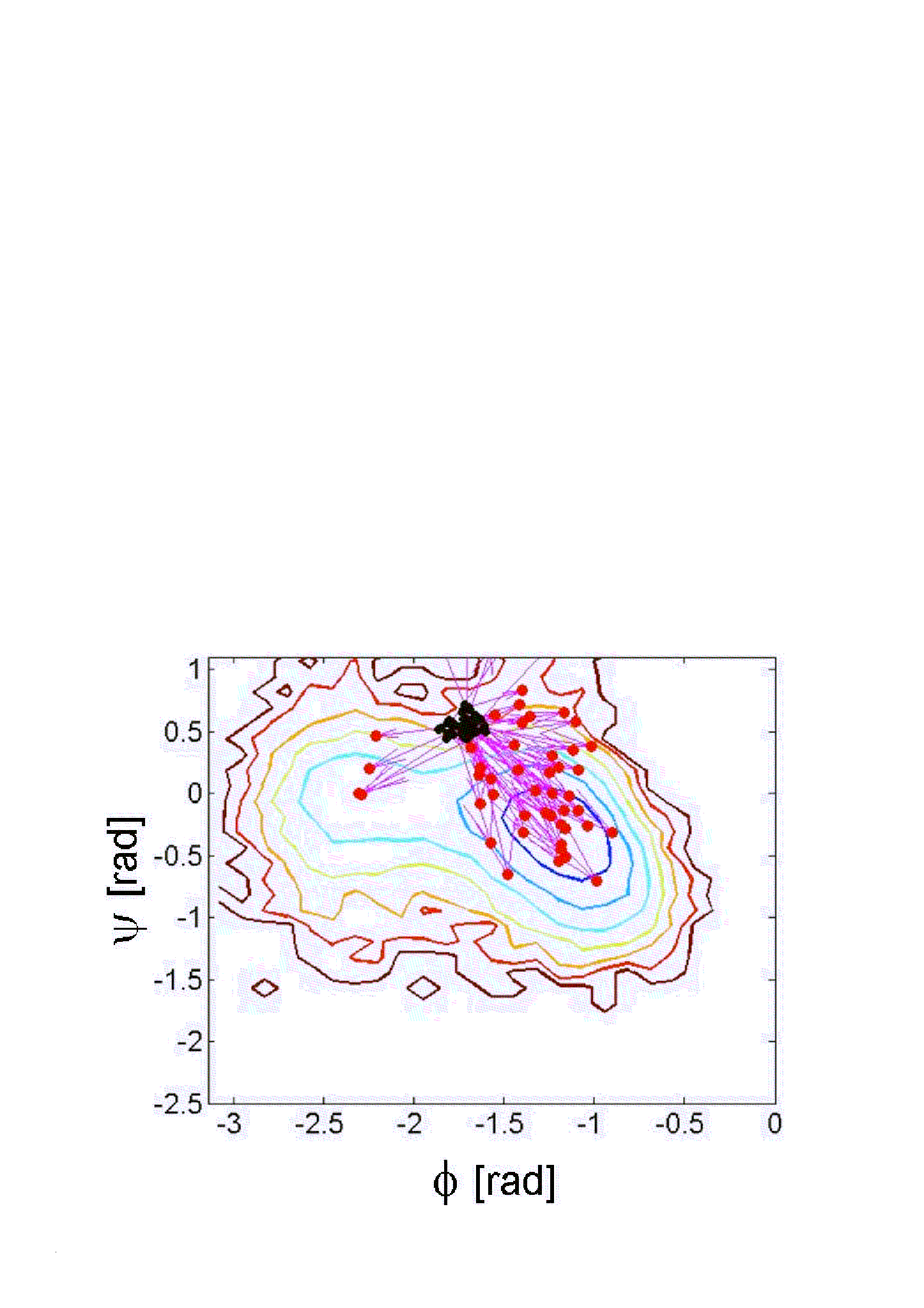}
\caption*{Figure 15}
\label{fig:aladip_extd}
\end{figure}

\clearpage
\newpage
\begin{figure}[htbp]
  \centering
  \includegraphics[bb=0 0 790 820,scale=0.6]{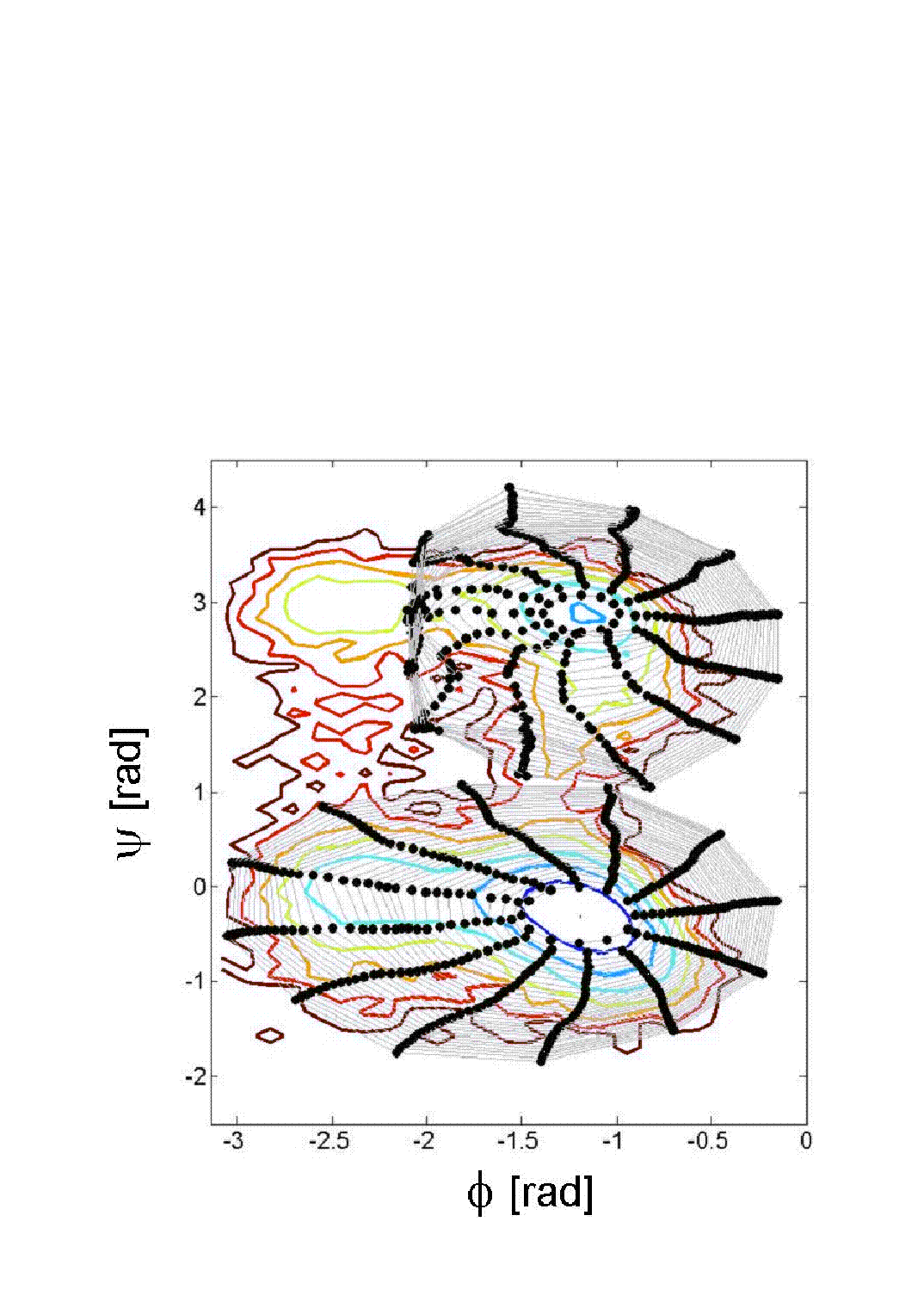}
\caption*{Figure 16}
\label{fig:aladip_ringall}
\end{figure}

\end{document}